\documentclass[journal,draftclsnofoot,onecolumn,12pt]{IEEEtran}
\usepackage{cite}
\ifCLASSINFOpdf
\else
\usepackage[dvips]{graphicx}
\fi
\usepackage{amssymb,bm,mathrsfs,bbm,amscd,amsmath}
\newtheorem{Proposition}{Proposition}
\newtheorem{Remark}{Remark}
\usepackage{paralist}
\usepackage[ruled]{algorithm2e}  

\SetKwInput{KwData}{\textbf{Initialization}}
\usepackage[aboveskip=5pt,belowskip=-15pt]{caption}
\usepackage{subfig}
\usepackage{verbatim}

\usepackage{color}

    \def\bm#1{\boldsymbol{#1}}

    \def\Complex{{\rm\rule[.23ex]{.03em}{1.1ex}\kern-.3em{C}}}

    \newcommand{\be}{\begin{equation}} \newcommand{\ee}{\end{equation}}
    \newcommand{\bea}{\begin{eqnarray}} \newcommand{\eea}{\end{eqnarray}}
    \newcommand{\benum}{\begin{enumerate}} \newcommand{\eenum}{\end{enumerate}}

    \newcommand{\qa}{{\bf a}}

    \newcommand{\qq}{{\bf q}}
    
    \newcommand{\qs}{{\bf s}}
    
    \newcommand{\qu}{{\bf u}}
    \newcommand{\qv}{{\bf v}}
    \newcommand{\qw}{{\bf w}}

    \newcommand{\qz}{{\bf z}}

    \newcommand{\qA}{{\bf A}}

    \newcommand{\qF}{{\bf F}}

    \newcommand{\qI}{{\bf I}}

    \newcommand{\qP}{{\bf P}}
    \newcommand{\qQ}{{\bf Q}}
    
    \newcommand{\qS}{{\bf S}}

    \newcommand{\qW}{{\bf W}}

    \newcommand{\qZ}{{\bf Z}}

    \newcommand{\qLambda}{{\boldsymbol \Lambda}}

    \newcommand{\tc}{\tilde{c}}

    \newcommand{\tq}{\tilde{q}}
    
    \newcommand{\ts}{\tilde{s}}

    \newcommand{\ty}{\tilde{y}}

    \newcommand{\tqs}{\tilde{\bf s}}

    \newcommand{\tqw}{\tilde{\bf w}}

    \newcommand{\tqQ}{\tilde{\bf Q}}
    
    \newcommand{\tqS}{\tilde{\bf S}}

    \newcommand{\bbC}{{\mathbb C}}

    \newcommand{\calF}{{\cal F}}
    \newcommand{\calG}{{\cal G}}

    \newcommand{\calL}{{\cal L}}

    \newcommand{\calO}{{\cal O}}
    \newcommand{\calP}{{\cal P}}
    
    \newcommand{\calR}{{\cal R}}

    \newcommand{\tr}{{\sf tr}}

    \newcommand{\Extr}{\mathop{\mathsf{Extr}}}

    \newcommand{\rmd}{{\rm d}}
    \newcommand{\rmD}{{\rm D}}

    \newcommand{\sfj}{{\sf j}}

\begin{document}
%
\title{Bayesian Optimal Data Detector for mmWave OFDM System with Low-Resolution ADC}
%
%
%

\author{Hanqing~Wang,~
        Chao-Kai~Wen,~
        and~Shi~Jin~
\thanks{H. Wang and S. Jin are with the National Mobile Communications Research Laboratory, Southeast University, Nanjing 210096, P. R. China. P (e-mail: $\rm hqwanglyt@seu.edu.cn;~jinshi@seu.edu.cn$).}
\thanks{C.-K. Wen is with the Institute of Communications Engineering, National Sun Yat-sen University, Kaohsiung, Taiwan (e-mail: $\rm chaokai.wen@mail.nsysu.edu.tw$).}
\thanks{Part of this work has been presented at IEEE ICCS 2016 in Shenzhen \cite{Wang-2016ICCS}.}
}

\maketitle

\begin{abstract}
Orthogonal frequency division multiplexing (OFDM) has been widely used in communication systems operating in the millimeter wave (mmWave) band to combat frequency-selective fading and achieve multi-Gbps transmissions, such as IEEE 802.15.3c and IEEE 802.11ad.
For mmWave systems with ultra high sampling rate requirements, the use of low-resolution analog-to-digital converters (ADCs) (i.e., 1--3 bits) ensures an acceptable level of power consumption and system costs.
However, orthogonality among subchannels in the OFDM system cannot be maintained because of the severe nonlinearity caused by low-resolution ADC, which renders the design of data detector challenging.
In this study, we develop an efficient algorithm for optimal data detection in the mmWave OFDM system with low-resolution ADCs.
The analytical performance of the proposed detector is derived and verified to achieve the fundamental limit of the Bayesian optimal design.
On the basis of the derived analytical expression, we further propose a power allocation (PA) scheme that seeks to minimize the average symbol error rate.
In addition to the optimal data detector, we also develop a feasible channel estimation method, which can provide high-quality channel state information without significant pilot overhead.
Simulation results confirm the accuracy of our analysis and illustrate that the performance of the proposed detector in conjunction with the proposed PA scheme is close to the optimal performance of the OFDM system with infinite-resolution ADC.
\end{abstract}
%


%
\IEEEpeerreviewmaketitle

\section{Introduction}

Millimeter wave (mmWave) communications utilize the spectrum range of 30 GHz to 300 GHz, where a large bandwidth is available, to achieve ultra high data rates \cite{RHeath2016JSTSP}.
Large-scale applications operating in the mmWave band are emerging, such as wireless local and personal area network systems \cite{r1,r0}, 5G cellular systems \cite{Rappaport-2013Access}, vehicular communications \cite{RHeath-2016FTN}, and wearables \cite{RHeath-2016Access}, because of this high rate supporting potential and the severe shortage of spectrum resource available in the sub-6 GHz bands.

Despite the potential advantage of high data rates, mmWave communications demand very high sampling frequencies on analog-to-digital converters (ADCs), where received analog signals are converted into digital signals for subsequent signal processing.
Unfortunately, the power consumption of an ADC unit increases quadratically with the sampling frequency and exponentially with the number of quantization bits at a sampling rate above 100 MSps \cite{Walden-1999JSAC,Murmann}.
Applying high speed (e.g., several GSps) and high precision (e.g., above 6 bits) ADCs at the mmWave receiver shall result in prohibitively high power consumption and system costs, particularly in mobile devices.
This issue is among the key bottlenecks in achieving mmWave systems.
A potential direction to pursue is the use of very-low-resolution ADCs (e.g., 1--3 bits\footnote{Current wireless communication systems typically equip 8--12 bit ADCs at their receivers.}) aligned with advanced signal processing techniques to mitigate the sacrifice in overall system performance \cite{Sundstrom2009TCSI}.
Several aspects of this direction have been investigated in the literatures,
including capacity analysis and capacity-achieving strategy for single-input single-output (SISO) channel \cite{Mezghani-08ISIT,Singh2009TCOM,Krone2010ITW} and multiple-input-multiple-output (MIMO) channel \cite{Mo-15TSP,Liang-16JSAC,Liang-16TCOM}, data detection for the MIMO system under frequency-flat channel \cite{Mezghani-10ISIT,Risi-2014ArXiv,Choi-16TCOM,TJiang2016TWC,CKWen2016TSP} and frequency-selective channel
\cite{Wang-15TWireless,Mollen-16TWireless,Studer2016TCOM,Nossek-2016ArXiv}, and channel estimation \cite{CKWen2016TSP,Mezghani-10WSA,Nossek-2016ArXiv,Li-16ArXiv,Mo-16ArXiv}.

Meanwhile, the signal transmitted over the mmWave channel, where the bandwidth is much wider than the coherence bandwidth, generally suffers from severe frequency-selective fading, which gives rise to serious inter-symbol interference (ISI). By adding a cyclic prefix (CP) for converting linear convolution into circular convolution and using the discrete Fourier transform (DFT), orthogonal frequency division multiplexing (OFDM) technology decomposes the ISI channel into a set of orthogonal subchannels with a bandwidth smaller than the coherence bandwidth \cite{Proakis2007}.

Consequently, the OFDM technology has been widely used in various wideband wireless communication systems to combat ISI caused by the frequency-selective fading.
In the mmWave range, standard systems, such as IEEE 802.11ad \cite{r1} and IEEE 802.15.3c \cite{r0}, operate in the 60 GHz band and use the OFDM technique to achieve data rates of up to multiple Gbps.

In this study, we focus on OFDM systems with low-resolution ADCs at the receiver.
We refer to such systems as quantized OFDM (Q-OFDM) systems.
The coarse quantization in the OFDM system causes strong nonlinear distortion on the received signals, such that the orthogonality among subchannels cannot be maintained in the Q-OFDM system and severe inter-carrier interference (ICI) occurs.
These issues render the design of data detection algorithms challenging because the simple one-tap equalizer used in conventional OFDM receivers can no longer perform well.
A traditional heuristic approximates the effect of hardware imperfections by using a linear model \cite{Matthaiou-15TWC}. These imperfections include phase-drifts, distortion noise, and amplified thermal noise. The additive quantization noise model (AQNM), which assumes that quantization noise is additive and independent, is a representative model of this method. This linear approximation facilitates the analysis of spectral efficiency and energy efficiency for systems with low-resolution ADCs, especially for massive MIMO systems \cite{LFan2015CL,Matthaiou-2016SAM,LFan-2016SAM}. Therefore, AQNM generates additional insights into system design perspective, such as the optimal number of base station (BS) antennas \cite{Matthaiou-2016SAM} as well as optimal pilot length \cite{LFan-2016SAM} and ADC resolution \cite{QBai-2013ISWCS}. However, AQNM cannot provide satisfactory approximation in the Q-OFDM system because this model completely ignores the ICI effect caused by the coarse ADC. Data detection based on the AQNM leads to significant performance loss, which will be confirmed by simulation results.

Although various studies on data detection problems, such as \cite{Mezghani-10ISIT,Risi-2014ArXiv,Choi-16TCOM,TJiang2016TWC,CKWen2016TSP}, have considered the exact quantization model, they are all dedicated to the data detection for general MIMO channels rather than for the Q-OFDM channels.
From the statistical inference perspective, very little difference exists between the Q-OFDM channel and the quantize MIMO channel in terms of data detection, which both involve inferring a random vector observed through a linear transformation followed by a nonlinear measurement channel.
However, the linear transformation matrix in the OFDM channel is orthogonal, whereas that in the quantize MIMO channel is independent and identically distributed (i.i.d.) random.

Furthermore, data detection algorithms proposed for the wideband channel \cite{Mollen-16TWireless,Studer2016TCOM,Nossek-2016ArXiv} are also sub-optimal for the Q-OFDM system.
The fast adaptive shrinkage/thresholding algorithm used in \cite{Studer2016TCOM} assumes that the transmitted symbols are drawn from a complex Gaussian distribution, which is not optimal for the detection of modulated signals. In \cite{Nossek-2016ArXiv}, an efficient data detection algorithm based on the generalized approximate message passing (GAMP) algorithm \cite{Rangan-2012Arxiv} was proposed.
GAMP is the most representative (and state-of-the-art) approach for the estimation of a random vector observed through a linear transformation followed by a componentwise, nonlinear measurement channel. However, GAMP has been proven optimal for i.i.d. waveforms only and not for the orthogonal waveform of our interest.
Moreover, the performance analysis of the GAMP-based detector is not available for the orthogonal waveform. Therefore, performing time-consuming Monte-Carlo simulations to evaluate the GAMP-based detector for the Q-OFDM system is inevitable.
Recent works in \cite{Ma-15SPL,Ma-15SPL-2,CKWen2016ISIT} revealed that the optimal inference for  i.i.d. transform matrices yields worse performance  for sparse signal recovery problems with orthogonal transform matrices.
Therefore, the detection performance under the Q-OFDM channel may be underestimated when employing the existing algorithms.

Thus far, the solution on how to achieve the best data detection performance for the Q-OFDM system is generally unknown.
This study takes the first step toward this direction. Specifically, we propose an optimal, computationally tractable data detector based on the Turbo iteration principle proposed in \cite{CKWen2016ISIT} and derive its corresponding state evolution (SE) equations. The uniqueness of this work is summarized as follows:

\begin{itemize}
	\item \emph{Optimality.}
	The SE equations of the proposed detector can match those of the Bayesian optimal detector derived via the replica theory.
	This indicates that the proposed detector can attain the optimal detection performance. Importantly, in contrast with direct computation of the Bayesian optimal solution, the proposed detector is computationally tractable.
	The symbol error rate (SER) of the proposed detector provides the lower bound for various detectors for the Q-OFDM system, which can be served as a benchmark for algorithm design and a foundation for evaluating the feasibility of utilizing low resolution ADCs in practical systems. We demonstrate through simulations that the proposed detector achieves better performance than the most representative GAMP-based detectors without any increase in computational complexity.
	
	\item \emph{Theoretical Analysability.}
	The SE analysis of the proposed algorithm is available.
	With SE analysis, performance metrics, such as the average SER, can be analytically determined without using time-consuming Monte Carlo simulations.
	Notably, the SE analysis demonstrates a decoupling principle, that is, the input-output relationship of the proposed detector on each subchannel can be \emph{decoupled} into a bank of equivalent additive white Gaussian noise (AWGN) channels.
	The decoupling principle enables the development of a power allocation (PA) algorithm to minimize the average SER across these equivalent AWGN channels.
	The simulations show that this PA scheme improves the SER performance significantly compared with the equal subchannel PA (ESPA).
	
	\item \emph{Flexibility.}
	The principle underlying the proposed detector provides a unified framework for solving a variety of detection and estimation problems.
	Under this unified framework, we also develop a feasible method for channel estimation to apply the proposed Q-OFDM detector to a practical scenario without the perfect CSI.
	The simulation results show that precise CSI can be acquired through the proposed scheme.
\end{itemize}

\emph{Notations.} This paper uses lowercase and uppercase boldface letters to represent vectors and matrices, respectively. For vector $\mathbf{a}$, the operator
$\mathrm{diag}(\mathbf{a})$ denotes the diagonal matrix with diagonal elements as the $\mathbf{a}$ entries. Moreover, the real and imaginary parts of a complex
scalar $a$ are represented by $a^R$ and $a^I$, respectively. The distribution of a proper complex Gaussian random variable $z$ with mean $\mu$ and variance $\nu$
is expressed as
\[z\sim\mathcal{CN}(z;\mu,\nu)=\frac{1}{\pi\nu}e^{-\frac{|z-\mu|^2}{\nu}}. \]
Similarly, $\mathcal{N}(z;\mu,\nu)$ denotes the probability density function (PDF) of a real Gaussian random variable $z$ with mean $\mu$ and variance $\nu$.
We let $\mathrm{D}z$ denote the real Gaussian integration measure
\[\mathrm{D}z=\phi(z)\mathrm{d}z~\text{with}~\phi(z)=\frac{1}{\sqrt{2\pi}}e^{-\frac{z^2}{2}}.\]
The cumulative Gaussian distribution function is defined as $\Phi(z)=\frac{1}{\sqrt{2\pi}}\int_{-\infty}^{z}e^{-\frac{t^2}{2}}\mathrm{d}t$, and the Q-function is defined as $Q(z)=1-\Phi(z)$.

\section{System Model}
We consider the OFDM system with $N$ orthogonal subchannels.
We let $\mathbf{s}=[s_1,s_2,\cdots,s_N]^{T}\in\mathcal{S}^N$ denote the input block to be transmitted in each subchannel, where $\mathcal{S}$ denotes the set of constellation points of the chosen modulation method, such as quadrature phase shift keying (QPSK) or quadrature amplitude modulation (QAM).
We allocate power $p_j$ to the $j$-th subchannel while keep the total power of the entire OFDM symbol constant to optimize some performance metrics, such as SER.
Specifically, the symbol in the $j$-th subchannel is multiplied by the scalar coefficient $\sqrt{p_j}$ and $\sum_{j=1}^{N} p_j = N\bar{P}$, where $\bar{P}$ is the average power per subchannel available in the transmitter. Then, we define a new diagonal
matrix $\mathbf{P}=\mathrm{diag}(p_1,p_2,\cdots,p_N)$. The frequency-domain block $\mathbf{P}^{\frac{1}{2}}\mathbf{s}$ is transformed to the time domain by the
$N$-point inverse DFT written as $\mathbf{F}^{H}\mathbf{P}^{\frac{1}{2}}\mathbf{s}$, where $\mathbf{F}$ denotes the normalized DFT matrix whose $(m,n)$-th entry is
$\frac{1}{{\sqrt N }} e^{-{2\pi j(n - 1)(m - 1)}/N}$.

The transmitted signal is filtered by a multipath channel, which can be represented by a tapped delay line model with $L$ taps. We let $g_i$ denote the
discrete-time impulse response of the $i$-th tag. The last $L_{\rm cp}$ $(L_{\rm cp}\geq L)$ time domain samples are appended as a CP at the beginning of each OFDM
symbol before transmitting it over the channel to avoid the ISI caused by the multipath channel. At the receiver, the analog signal is discretized after
down-converting the received signal into the analog baseband. After CP removal, the (unquantized) received block of OFDM symbol can be written as
\begin{equation}\label{eII_01}
	\mathbf{y}=\mathbf{G}\mathbf{F}^H\mathbf{P}^{\frac{1}{2}}\mathbf{s}+\mathbf{n},
\end{equation}
where $\mathbf{G}\in\mathcal{C}^{N\times N}$ is the circulant matrix with $\mathbf{g} = [g_1, \, g_2, \cdots, g_N]^T$ being its first column and $g_j=0$ for
$(L+1)\leq j\leq N$, and $\mathbf{n}$ is the AWGN vector with zero mean and covariance matrix $\sigma^2\mathbf{I}$. The circulant matrix $\mathbf{G}$ can be
decomposed as
\begin{equation}\label{eII_03}
	\mathbf{G}=\mathbf{F}^{H}\mathrm{diag}(\mathbf{h})\mathbf{F},
\end{equation}
where $\mathbf{h}$ denotes the frequency-domain channel presentation obtained by operating DFT on the first column of $\mathbf{G}$, that is,
$\mathbf{h}=\mathbf{F}\mathbf{g} $. Substituting \eqref{eII_03} into \eqref{eII_01}, we can rewrite \eqref{eII_01} as
\begin{equation}\label{eII_04}
	\mathbf{y}=\mathbf{F}^H\mathrm{diag}(\mathbf{h}')\mathbf{s}+\mathbf{n}.
\end{equation}
where $\mathbf{h}'$ denotes the channel vector comprised of the diagonal entries of the matrix $\mathrm{diag}(\mathbf{h})\mathbf{P}^{\frac{1}{2}}$, that
is, $\mathbf{h}'=[\sqrt{p_1}h_1,\sqrt{p_2}h_2,\cdots,\sqrt{p_N}h_N]^T$.

Each element $y_j$ of the received signal $\mathbf{y}$ is quantized using a complex-valued quantizer $\mathcal{Q}_c(\cdot)$, which consists of two real-valued
quantizers $\mathcal{Q}(\cdot)$ that quantize the real and imaginary parts of $y_j$ separately and independently, that is,
\begin{equation}\label{eII_08}
	q_j=\mathcal{Q}_c(y_j)=\mathcal{Q}(y^R_j)+j\mathcal{Q}(y^I_j).
\end{equation}
We consider $\mathcal{Q}(\cdot)$ as a $B$-bit quantizer, which maps the real-valued input $y^R_j$ or $y^I_j$ to one of the $2^B$ discrete values. The
output is specifically assigned the value $c_b$, that is, the $b$-th discrete value, when the quantizer input is within the interval $(r_{b-1},r_{b}]$, where
$-\infty=r_{0}<r_1<\cdots<r_{2^{B-1}}<r_{2^B}=\infty$ are the thresholds. We take $c_b$ as the centroid of the interval $(r_{b-1},r_{b}]$. The quantized received
signal can be denoted as
\begin{equation}\label{eII_07}
	\mathbf{q}=\mathcal{Q}_c(\mathbf{F}^H\mathrm{diag}(\mathbf{h}')\mathbf{s}+\mathbf{n}).
\end{equation}

Data detection aims to recover the transmitted symbol $\mathbf{s}$ from the quantized signal $\mathbf{q}$ given by the linear mixing model \eqref{eII_07} with linear transformation matrix $\mathbf{F}^H$. The conventional OFDM receiver performs DFT directly
on the quantized signal $\mathbf{q}$ and yields $\tilde{\mathbf{q}}=\mathbf{F}\mathbf{q}$. The decision rule follows the one-tap equalizer given by
\begin{equation} \label{eq:convML}
	\hat{s}_j=\mathop{{\rm { argmin}}}\limits_{s\in\mathcal{S}} \left|\frac{\tilde{q}_j}{h'_j}-s \right|^2,  ~\mbox{for}~j=1,2,\cdots,N.
\end{equation}
For infinite-precision quantization, that is, $\mathbf{q}=\mathbf{y}$, the DFT operation on $\mathbf{q}$ enables the signal at each subchannel to be an AWGN
observation of the product of the transmitted symbol and its corresponding frequency-domain channel response. Therefore, \eqref{eq:convML} is the optimal decision rule based on the maximum likelihood (ML) criteria. However, this conventional OFDM detector, which employs a one-tap equalizer, is no longer optimal for the low-resolution quantization case in which the
orthogonality among subchannels is not preserved.

\begin{Remark}
	Beamforming techniques operating in the RF domain shall be used at the transmitter and receiver to overcome the high propagation loss in mmWave band.
	Markedly, \eqref{eII_04} is a concise equivalent representation for the input-output relationship of the mmWave OFDM system using analog transmitter and receiver beamforming
	with one transmitted and one received data stream as depicted in \cite[Fig. 2]{RHeath2016JSTSP}. Specifically, in this system, each element of $\mathbf{h}$ is
	expressed as \cite{Yin-13GLCOM}
	\begin{equation}
		h_j = \mathbf{w}^{\rm RX}\mathbf{H}_j\mathbf{w}^{\rm TX},
	\end{equation}
	where $\mathbf{w}^{\rm RX}\in\mathcal{C}^{1\times N_R}$ and $\mathbf{w}^{\rm TX}\in\mathcal{C}^{N_T\times 1}$ are beamforming vectors at the receiver and transmitter, respectively; 
	$\mathbf{H}_j\in\mathcal{C}^{N_R\times N_T}$ represents the channel response matrix at the $j$-th subchannel; and $N_T$ and $N_R$ are the number of
	transmit and receive antennas, respectively. \hfill\ensuremath{\blacksquare}
\end{Remark}

\begin{Remark}
	As the key technology for the next generation mobile communications, mmWave communications aligned with large-scale antenna array are definitely exploited for multi-stream and multi-user scenarios. Although designed for single stream problem, the proposed algorithm can be also employed by the receiver of uplink transmission of the cellular systems (i.e., the BS). For uplink transmission, the analog beamforming is implemented for the spatial division of different users.
	In addition, narrow beam is steered by the analog beamforming to form high-directional spatial links between different users and the BS. Therefore, following proper user selection, the entire uplink transmission can be approximately decomposed into several parallel single-stream communications, and the proposed detector can be employed for the optimal detection at the BS side of each individual spatial link. The proposed detector is advantageous considering that digital beamforming can be further applied to multiuser interference mitigation. However, the topic is beyond the scope of this paper and thus left for future work.\hfill\ensuremath{\blacksquare}
\end{Remark}

\section{Optimal Data Detection}

In this section, we explain the theoretical foundation for Bayesian inference and introduce the data detection algorithm. We first assume that the perfect channel
state information at the receiver (CSIR) $\mathbf{h}'$ is available to elucidate the concept. The performance analysis, the PA scheme, and the channel
estimation method will be introduced in Section IV and V.

\subsection{Theoretical Foundation}

Before proceeding, we define two auxiliary vectors
\begin{equation} \label{eq:def_x&z}
	\mathbf{x}=\mathrm{diag}(\mathbf{h}')\mathbf{s},~~~ \mathbf{z}=\mathbf{F}^H\mathbf{x}
\end{equation}
to facilitate our subsequent discussion. And we specify the likelihood function, which plays a key role in Bayesian inference. With the perfect CSIR $\mathbf{h}'$, the likelihood function is the distribution of the quantized signal
$\mathbf{q}$ conditioned on the transmitted vector $\mathbf{s}$. From \eqref{eII_07}, it can be given by
\begin{equation}\label{eII_10}
	\mathrm{P}(\mathbf{q} \mid \mathbf{s};\mathbf{h}') = \prod_{j=1}^{N} \mathrm{P}_{\mathrm{out}}(q_j\mid z_j).
\end{equation}
The factorization of $\mathrm{P}(\mathbf{q} \mid \mathbf{s};\mathbf{h}')$ is derived from the fact that from \eqref{eII_07}, the value of $q_i$ given $z_i$ depends only on $n_i$, and the elements of AWGN vector $\mathbf{n}$ are statistically independent. According to the property of the complex-valued quantizer \eqref{eII_08}, we derive that
\begin{equation} \label{eq:P_outEach}
	\mathrm{P}_{\mathrm{out}}(q_j\mid z_j)=\mathrm{P}(q^R_j\mid z^R_j)\mathrm{P}(q^I_j\mid z^I_j),
\end{equation}
where $\mathrm{P}(q^R_j\mid z^R_j)$ denotes the probability of observing the real part quantized output $q^R_j$ given the real part of noiseless unquantized received signal $z^R_j$. Specifically,
\begin{equation}\label{eII_09}
	\mathrm{P}(q^R_j\mid z^R_j)= \Phi{\left(\frac{\sqrt{2}(z^R_j-l(q^R_j))}{\sigma}\right)} \hspace{-0.05cm} - \hspace{-0.05cm} \Phi{\left(\frac{\sqrt{2}(z^R_j-u(q^R_j))}{\sigma}\right)}
\end{equation}
where $l(q^R_j)$ and $u(q^R_j)$ denote the corresponding lower and upper bounds of the quantizer output value $q^R_j$. For example, when $q^R_j=c_b$,
$l(q^R_j)=r_{b-1}$ and $u(q^R_j)=r_{b}$. The corresponding probability for the imaginary part $\mathrm{P}(q^I_j\mid z^I_j)$ can be given analogously.

According to the Bayesian rule, the posterior probability can be obtained by
\begin{equation} \label{eq:postP}
	{\mathrm{P}(\mathbf{s} \mid \mathbf{q};\mathbf{h}')}
	= \frac{ \mathrm{P}(\mathbf{q} \mid \mathbf{s};\mathbf{h}') \mathrm{P}(\mathbf{s}) }{ \mathrm{P}(\mathbf{q} ;\mathbf{h}')},
\end{equation}
where $\mathrm{P}(\mathbf{q} \mid \mathbf{s};\mathbf{h}')$ is the likelihood function defined in \eqref{eII_10}, $\mathrm{P}(\mathbf{s})$ is the prior
distribution, and $\mathrm{P}(\mathbf{q} ;\mathbf{h}')$ is the marginal distribution computed by
\begin{equation}\label{eq:MQ}
	\mathrm{P}(\mathbf{q} ;\mathbf{h}')=\int_{\mathbf{s}}\mathrm{P}(\mathbf{q} \mid \mathbf{s};\mathbf{h}') \mathrm{P}(\mathbf{s}){\rm d} \mathbf{s}.
\end{equation}

In this paper, we consider that the elements of $\mathbf{s}$ are i.i.d., therefore
\begin{equation} \label{eq:proX}
	\mathrm{P}(\mathbf{s}) = \prod_{j=1}^{N} \mathrm{P}(s_j),
\end{equation}
and $s_j$'s are drawn from a set of constellation points with equal probabilities, thus $\mathrm{P}(s_j) = 1/|\mathcal{S}|$ for $s_j \in \mathcal{S}$.

Using the posterior probability (\ref{eq:postP}), the marginal posterior probability can be obtained via
\begin{equation} \label{eq:postP_sj}
	\mathrm{P}(s_j \mid \mathbf{q};\mathbf{h}') = \int_{\mathbf{s} \setminus s_j} \mathrm{P}(\mathbf{s} \mid \mathbf{q};\mathbf{h}') {\rm d} \mathbf{s}.
\end{equation}
The posterior mean achieves the \emph{minimum mean-square error} (MMSE), and its $j$-th element can be expressed as:
\begin{equation}  \label{eq:postMean_x}
	\bar{s}_j = \mathrm{E}\left[ s_j \mid \mathbf{q};\mathbf{h}' \right] = \int s_j  \mathrm{P}(s_j \mid \mathbf{q};\mathbf{h}') {\rm d} s_j .
\end{equation}
Moreover, the widely used \emph{maximum a posterior} (MAP) decision rule is given by
\begin{equation}\label{eq:eqi_ML}
	\hat{s}_j=  \mathop{{\rm argmax}}\limits_{s\in\mathcal{S}}  \mathrm{P}(s_j \mid \mathbf{q};\mathbf{h}') .
\end{equation}

The Bayesian MMSE estimation (\ref{eq:postMean_x}) and MAP inference (\ref{eq:eqi_ML}) are computationally intractable in this case because the calculation of marginal
posterior probability in (\ref{eq:postP_sj}) involves the high-dimensional integral. We resort to a recently developed approximation technique called the
generalized Turbo (GTurbo) principle \cite{CKWen2016ISIT} to calculate the posterior mean \eqref{eq:postMean_x}  iteratively. We demonstrate the adoption of the
GTurbo principle for data detection in the subsequent subsection.

\begin{Remark}
	The posterior probability \eqref{eq:postP} together with the likelihood \eqref{eII_10} and the prior \eqref{eq:proX} can be represented as a graphical model \cite{Koller-BOOK09} with the elements of $\mathbf{s}$ and $\mathbf{q}$ being its variable nodes and factor nodes respectively.
	Belief propagation (BP) is a typical technique for calculating marginal distributions and can often provide good approximations for margins on sparse graphical models.
	However, \eqref{eII_07} corresponds to a dense graphical model where each factor node interacts with all variable nodes because of the linear transformation $\mathbf{F}^H$.
	GAMP \cite{Rangan-2012Arxiv} is an approximate version of BP that emerges recently and demonstrates good performance in dense graphical models.
	A closely related work \cite{Nossek-2016ArXiv} investigates the same data detection problem as in this
	study using GAMP. However, GAMP was proven to yield the optimal solutions to (\ref{eq:postMean_x}) and (\ref{eq:eqi_ML}) only if the entries of linear transformation matrix of the linear mixing model \eqref{eII_07} are independent.
	The superiority of the proposed algorithm based on the GTurbo principle over the existing algorithms will be shown through simulation results. \hfill\ensuremath{\blacksquare}
\end{Remark}

\subsection{GTurbo-based Algorithm}

\begin{figure*}[!h]
    \centering
    \includegraphics[scale=0.8]{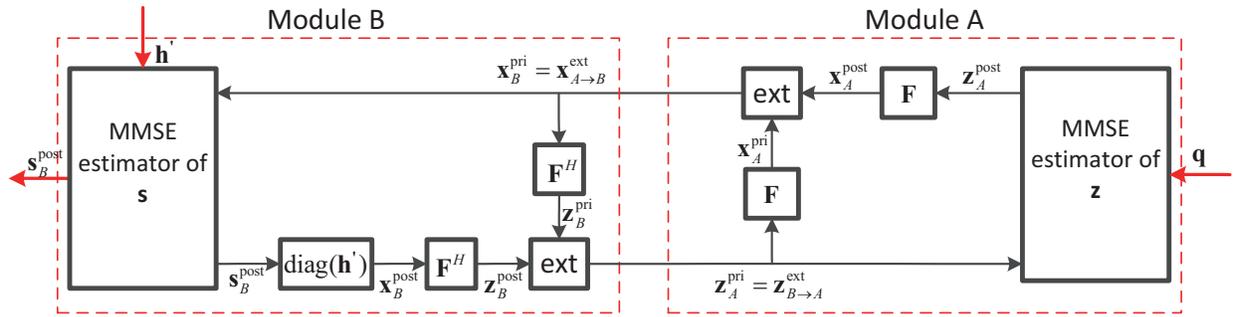}
    \caption{ The block diagram of GTurbo-based data detection algorithm with the perfect CSIR. The ``ext'' block represents the extrinsic information computation.
        The block of a certain matrix represents the left-multiplying the input vector by the matrix in the block.\label{F1}}
\end{figure*}

The GTurbo-based data detection algorithm for the Q-OFDM system is presented in Algorithm~\ref{A1}, and the corresponding block diagram is illustrated in Fig.~\ref{F1}. This
algorithm comprises of two modules: Module A produces the direct coarse estimation of $\mathbf{x}$ from the relationship ${{\mathbf{x}}={\mathbf{F}}{\mathbf{z}}}$
in \eqref{eq:def_x&z} without considering prior $\mathrm{P}(\mathbf{s})$, whereas Module B refines the estimate by considering prior $\mathrm{P}(\mathbf{s})$. The
two modules are executed iteratively until convergence.

\begin{algorithm}[!ht]
    \caption{GTurbo-based Data Detection with Perfect CSI\label{A1}}
    \small
    \KwData{$\mathbf{z}_A^{{\rm pri}}=\mathbf{0}_{N\times 1}$, $v_A^{{\rm pri}}=\frac{1}{N}\sum_{j=1}^{N}|h'_j|^2$\;}
    \For{$t=1:T_{max}$}{
        \textbf{Module A:}\\
        (1) Compute the posteriori mean/variance of $\mathbf{z}$:
        \begin{subequations} \label{eA1}
            \begin{align}
            &z_{j,A}^{\rm post}=\mathrm{E}\left[z^R_j\mid q^R_j\right]+j\mathrm{E}\left[z^I_j\mid q^I_j\right],\label{eA1_01}\\
            &v_{j,A}^{\rm post}=\mathrm{var}\left[z^R_j\mid q^R_j\right]+\mathrm{var}\left[z^I_j\mid q^I_j\right],\label{eA1_02}
            \end{align}
        \end{subequations}
        (2) Compute the extrinsic mean/variance of $\mathbf{x}$:
        \begin{subequations} \label{eq:eA1}
            \begin{align}
            &v_{A}^{\rm post}=\frac{1}{N}\sum\limits_{j=1}^{N}v_{j,A}^{\rm post},\label{eA1_04}\\
            &v_B^{{\rm pri}}=v_{A}^{\rm ext}=\left(\frac{1}{v_{A}^{\rm post}}-\frac{1}{v_A^{{\rm pri}}}\right)^{-1},\label{eA1_05}\\
            &\mathbf{x}_B^{{\rm pri}}=\mathbf{x}_{A}^{\rm ext}=v_{A}^{\rm ext}\left(\frac{\mathbf{F}\mathbf{z}_A^{\rm post}}{v_{A}^{\rm post}}-\frac{\mathbf{F}\mathbf{z}_A^{{\rm pri}}}{v_A^{{\rm pri}}}\right)\label{eA1_06},
            \end{align}
        \end{subequations}
        \textbf{Module B:}\\
        (3) Compute the posteriori mean/variance of $\mathbf{s}$:
        \begin{subequations}
            \begin{align}
            &s_{j,B}^{\rm post}=\mathrm{E}\left[s_{j}\mid h'_{j}, x_{j,B}^{{\rm pri}}\right],\label{eA1_07}\\
            &v_{j,B}^{\rm post}=\mathrm{var}\left[s_{j}\mid h'_{j}, x_{j,B}^{{\rm pri}}\right],\label{eA1_08}
            \end{align}
        \end{subequations}
        (4) Compute the extrinsic mean/variance of $\mathbf{z}$:
        \begin{subequations} \label{eq:e0530}
            \begin{align}
            &x_{j,B}^{\rm post} = {h'_j}s_{j,B}^{\rm post},\label{e0530_07a}\\
            &v_{B}^{\rm post}=\frac{1}{N}\sum\limits_{j=1}^{N}|h'_j|^2v_{j,B}^{\rm post}\label{e0530_07b},\\
            &v_A^{{\rm pri}}=v_{B}^{\rm ext}=\left(\frac{1}{v_{B}^{\rm post}}-\frac{1}{v_B^{{\rm pri}}}\right)^{-1},\label{e0411_08a}\\
            &\mathbf{z}_A^{{\rm pri}}=\mathbf{z}_{B}^{\rm ext}=v_{B}^{\rm ext}\left(\frac{\mathbf{F}^H\mathbf{x}_B^{\rm post}}{v_{B}^{\rm post}}-\frac{\mathbf{F}^H\mathbf{x}_B^{{\rm pri}}}{v_B^{{\rm pri}}}\right). \label{e0411_08b}
            \end{align}
        \end{subequations}
    }
\end{algorithm}

We provide a number of detailed explanations for Algorithm~\ref{A1}. In Module A, $\mathbf{z}_A^{{\rm post}}$ can be viewed as the Bayesian MMSE estimation of
$\mathbf{z}$ from the relationship
\begin{equation}\label{eq:AWGN_Z}
	\mathbf{q} = {\cal Q}_{c} {\left( \mathbf{z} + \mathbf{n} \right)}, ~\mbox{and}~~
	\mathbf{z} = \mathbf{z}_A^{\rm pri} + \bm{\omega}_A,
\end{equation}
where $\bm{\omega}_A\sim\mathcal{CN}(\mathbf{0},v_A^{{\rm pri}}\bf{I})$. Specifically, \eqref{eA1_01} and \eqref{eA1_02} compute the posteriori mean and variance of $z_j$ respectively, given its corresponding quantized observation $q_j$, where $\mathrm{E}\left[z^R_j\mid q^R_j\right]$ and $\mathrm{var}\left[z^R_j\mid q^R_j\right]$ denote the expectation and variance of $z^R_j$ with respect to (w.r.t.) the posterior probability
\[\mathrm{P}(z^R_j\mid q^R_j)=\frac{\mathrm{P}(q^R_j\mid z^R_j)\mathrm{P}(z^R_j)}{\int_{-\infty}^{\infty}\mathrm{P}(q^R_j\mid z^R_j)\mathrm{P}(z^R_j)\mathrm{d}z^R_j},\]
where $\mathrm{P}(q^R_j\mid z^R_j)$ is given by \eqref{eII_09}, and $\mathrm{P}(z^R_j)=\mathcal{N}(z^R_j;z_{j,A}^{{\rm pri},R},\frac{1}{2}v_A^{{\rm pri}})$ for the given $v_A^{{\rm pri}}$ and $\mathbf{z}_A^{\rm pri}$ under the assumption \eqref{eq:AWGN_Z}.
Following the derivation of \cite[Appendix A]{CKWen2016TSP}, the explicit expressions of the posteriori mean and variance of $z^R_j$ given $q^R_j$ can be obtained by
\begin{subequations}
	\begin{align}
		&\mathrm{E}\left[z^R_j\mid q^R_j\right]=z_{j,A}^{{\rm pri},R}+\frac{v_A^{{\rm pri}}}{\sqrt{2(v_A^{{\rm pri}}+\sigma^2)}}\left(\frac{\phi(\eta_1)-\phi(\eta_2)}{\Phi(\eta_1)-\Phi(\eta_2)}\right),\label{eIII_01}\displaybreak[0]\\
       &\mathrm{var}\left[z^R_j\mid q^R_j\right]=\frac{v_A^{{\rm pri}}}{2}-\frac{(v_A^{{\rm pri}})^2}{2(v_A^{{\rm pri}}+\sigma^2)}\times\left[\left(\frac{\phi(\eta_1)-\phi(\eta_2)}{\Phi(\eta_1)-\Phi(\eta_2)}\right)^2+\frac{\eta_1\phi(\eta_1)-\eta_2\phi(\eta_2)}{\Phi(\eta_1)-\Phi(\eta_2)}\right], \label{eIII_02}
	\end{align}
\end{subequations}
where
\begin{equation}\label{eIII_03}
	\eta_1=\frac{z_{j,A}^{{\rm pri},R}-u(q^R_j)}{\sqrt{(v_A^{{\rm pri}}+\sigma^2)/2}},~~\eta_2=\frac{z_{j,A}^{{\rm pri},R}-l(q^R_j)}{\sqrt{(v_A^{{\rm pri}}+\sigma^2)/2}}.
\end{equation}
Furthermore, $\mathrm{E}\left[z^I_j\mid q^I_j\right]$ and $\mathrm{var}\left[z^I_j\mid q^I_j\right]$ can be computed analogously by replacing $z_{j,A}^{{\rm
		pri},R}$ with $z_{j,A}^{{\rm pri},I}$ in the computation for $\eta_1$ and $\eta_2$ in \eqref{eIII_03}.

From \eqref{eq:def_x&z}, we derive ${{\mathbf{x}}={\mathbf{F}}{\mathbf{z}}}$. Therefore, the posteriori mean and variance of ${\mathbf{x}}$ can be computed by
${\mathbf{F}}\mathbf{z}_{A}^{\rm post}$ and ${\mathbf{F}}\mathrm{diag}(v_{1,A}^{\rm post},v_{2,A}^{\rm post},\cdots,v_{N,A}^{\rm post}){\mathbf{F}}^{H}$,
respectively. To reduce the computational complexity, we replace $\mathrm{diag}(v_{1,A}^{\rm post},v_{2,A}^{\rm post},\cdots,v_{N,A}^{\rm post})$ with $(\frac{1}{N}\sum_{j=1}^{N}v_{j,A}^{\rm
	post})\mathrm{\mathbf{I}}$ as in \eqref{eA1_04}. Subsequently, the extrinsic mean and variance of $\mathbf{x}$ are computed
by \eqref{eA1_05} and \eqref{eA1_06} similar to the concise formulas in \cite[(14) and (15)]{QGuo2011CL}, which are then used as the inputs $v_B^{{\rm pri}}$ and
$\mathbf{x}_B^{{\rm pri}}$ of Module B. Therefore, Module A produces an estimate of $(\mathbf{x},\mathbf{z})$ in which $\mathbf{x}$ is estimated through the linear
relation \eqref{eq:def_x&z} without considering prior $\mathrm{P}(\mathbf{s})$, whereas $\mathbf{z}$ is the Bayesian MMSE estimation by considering the likelihood
$\mathrm{P}(\mathbf{q} \mid \mathbf{z})$.

Subsequently, we turn to the MMSE estimation of $\mathbf{s}$ processed in Module B. Initially, $\mathbf{x}_B^{{\rm pri}}$ is assumed as an AWGN observation of
$\mathbf{x}=\mathrm{diag}(\mathbf{h}')\mathbf{s}$, that is,
\begin{equation}\label{eIII_04}
	\mathbf{x}_B^{{\rm pri}}=\mathrm{diag}(\mathbf{h}')\mathbf{s}+\bm{\omega}_B,
\end{equation}
where $\bm{\omega}_B\sim\mathcal{CN}(\mathbf{0},v_B^{{\rm pri}}\bf{I})$. Using the aforementioned assumption and the given frequency-domain channel response
$\mathbf{h}'$, we compute the posteriori mean and variance of $\mathbf{s}$ in \eqref{eA1_07} and \eqref{eA1_08} taken w.r.t. the posterior probability
distribution
\begin{equation}\label{e1019_01}
	\mathrm{P}(s_{j}\mid x_{j,B}^{{\rm pri}}; h'_{j})=\frac{\mathcal{CN}(x_{j,B}^{{\rm pri}};h'_js_j,v_B^{{\rm pri}})\mathrm{P}(s_j)}{ \sum\limits_{ s\in\mathcal{S} } \mathcal{CN}(x_{j,B}^{{\rm pri}};h'_js,v_B^{{\rm pri}})\mathrm{P}(s) }.
\end{equation}
Consequently, the explicit expressions of $s_{j,B}^{\rm post}$ and $v_{j,B}^{\rm post}$ can be derived as
\begin{subequations}
	\begin{align}
		&s_{j,B}^{\rm post} = \frac{{\sum\limits_{ s\in\mathcal{S} } {s\, \mathcal{CN}\left(s;\frac{{x_{j,B}^{{\rm pri}}}}{{{h'_j}}},\frac{{v_B^{{\rm pri}}}}{{{{\left| {{h'_j}} \right|}^2}}}\right)} }}{{ \sum\limits_{ s\in\mathcal{S} } { \mathcal{CN}\left(s;\frac{{x_{j,B}^{{\rm pri}}}}{{{h'_j}}},\frac{{v_B^{{\rm pri}}}}{{{{\left| {{h'_j}} \right|}^2}}}\right)} }},\displaybreak[0]\label{eIII_05}\\
		&v_{j,B}^{\rm post} = \frac{{\sum\limits_{ s\in\mathcal{S} } { {{\left| {s} \right|}^2}\mathcal{CN}\left(s;\frac{{x_{j,B}^{{\rm pri}}}}{{{h'_j}}},\frac{{v_B^{{\rm pri}}}}{{{{\left| {{h'_j}} \right|}^2}}}\right)} }}{{ \sum\limits_{ s\in\mathcal{S} } { \mathcal{CN}\left(s;\frac{{x_{j,B}^{{\rm pri}}}}{{{h'_j}}},\frac{{v_B^{{\rm pri}}}}{{{{\left| {{h'_j}} \right|}^2}}}\right)} }} - {\left| {s_{j,B}^{\rm post}} \right|^2}.\label{eIII_06}
	\end{align}
\end{subequations}

Similar to those in \eqref{eq:eA1}, $\mathbf{z}_B^{{\rm post}}$ is estimated directly based on the relationship ${{\mathbf{z}}={\mathbf{F}}^H{\mathbf{x}}}$. Then, the extrinsic
mean and variance of $\mathbf{z}$ are evaluated in \eqref{e0411_08a} and \eqref{e0411_08b}, respectively. Therefore, Module B produces an estimate of
$(\mathbf{s},\mathbf{z})$ in which $\mathbf{s}$ is the Bayesian MMSE estimation by considering prior $\mathrm{P}(\mathbf{s})$, whereas $\mathbf{z}$ is estimated
through the linear relation \eqref{eq:def_x&z} without considering the likelihood $\mathrm{P}(\mathbf{q} \mid \mathbf{z} )$.

Algorithm~\ref{A1} aims to calculate the marginal posterior probability in an iterative manner. After the convergence of the iteration, we obtain the estimated
marginal posterior probability $\mathrm{P}(s_{j}\mid x_{j,B}^{{\rm pri}}; h'_{j})=\mathcal{CN}(s_j;s_{j,B}^{\rm post},v_{j,B}^{\rm post})$. Thus, the posterior mean in \eqref{eq:postMean_x} is obtained as $s_{j,B}^{\rm
	post}$, and the MAP inference in \eqref{eq:eqi_ML} is equivalent to find $s\in\mathcal{S}$ with the shortest distance to $s_{j,B}^{\rm post}$, that is,
\begin{equation}\label{eq:asy_ML}
	\hat{s}_j= \mathop{{\rm argmin}}\limits_{s\in\mathcal{S}} \left|s-s_{j,B}^{\rm post}\right|^2.
\end{equation}

\section{State Evolution and Power Allocation}

The asymptotic performance of the proposed algorithm can be characterized by the recursion of a set of SE equations \cite{CKWen2016ISIT}. We derive these equations
in the large-system limit where $N\to\infty$ in Section IV-A. Subsequently, we show the decoupling principle and develop a subchannel power allocation scheme to
minimize the SER in Section IV-B.
Finally in Section IV-C, we analyze the complexity of the proposed algorithms.

\subsection{State Evolution}

From the explanations introduced in Section III-B, we observe that the performance of the detector is determined by $v_B^{{\rm pri}}$, which can be viewed as the
average noise power of the equivalent AWGN channels in \eqref{eIII_04}. In addition, $v_A^{{\rm pri}}$ and $v_B^{{\rm pri}}$ are mutually dependent in a recursive
manner as shown in \eqref{eA1_05} and \eqref{e0411_08a}, respectively. Therefore, we define the following two states to characterize the performance of the
detector:
\begin{equation}\label{eIII_08}
	\eta \triangleq \frac{1}{v_B^{{\rm pri}}},~\text{and}~\nu \triangleq v_A^{{\rm pri}}.
\end{equation}
In addition, we define the MMSE of $s$ given its AWGN observation $r = s + \omega$ as
\[\mathrm{mmse}(\eta) \triangleq \mathrm{E}[|s-\mathrm{E}[s|r]|^2],\]
where $\omega\sim\mathcal{CN}(0,\eta^{-1})$, the outer expectation is taken w.r.t. the distribution $\mathrm{P}(s)$, whereas the inner expectation is taken w.r.t.
the marginal distribution $\int\mathrm{P}(r|s)\mathrm{P}(s)\mathrm{d}s$. For example, if $s$ is drawn from the equiprobable QPSK constellation, then
$\mathrm{mmse}(\eta)$ can be derived as \cite{CKWen2016TSP}
\begin{equation}
	\mathrm{mmse}(\eta)=1- \int \tanh(\eta+\sqrt{\eta}z)\mathrm{D}z.
\end{equation}

By evaluating the two states in the large-system limit, Proposition \ref{Pro:SE} can be derived. The calculation details are provided in Appendix \ref{AP1}.
\begin{Proposition} \label{Pro:SE}
	In the large-system limit, the SE of Algorithm~\ref{A1} can be characterized by
	\begin{subequations} \label{eq:SE_parameters}
		\begin{align} &\vartheta^{t}=\frac{1}{2}\sum\limits_{b=1}^{2^B}\int_{-\infty}^{\infty}\frac{\left[\Psi'\left(c_b;\sqrt{\frac{v_x-\nu^{t}}{2}}z,\frac{\sigma^2+\nu^{t}}{2}\right)\right]^2}{\Psi\left(c_b;\sqrt{\frac{v_x-\nu^{t}}{2}}z,\frac{\sigma^2+\nu^{t}}{2}\right)}\mathrm{D}z,\displaybreak[0]\label{eq:SE1}\\
			&\eta^{t+1}=\frac{1}{(\vartheta^{t})^{-1}-\nu^t},\displaybreak[0]\label{eq:SE2}\\
			&\nu^{t+1}=\left(\frac{1}{\frac{1}{N}\sum\limits_{j=1}^{N}|h'_j|^2\mathrm{mmse}(|h'_j|^2\eta^{t+1})}-\eta^{t+1}\right)^{-1},\label{eq:SE3}
		\end{align}
	\end{subequations}
	where $t$ denotes the iteration index, the initialization $\upsilon^0=v_x\triangleq\frac{1}{N}\sum_{j=1}^{N}|h_j|^2p_j$, and
	\begin{align}
		&\Psi\left(c_b;z,u^2\right)\triangleq\Phi\left(\frac{z-r_{b-1}}{u}\right)-\Phi\left(\frac{z-r_{b}}{u}\right),\displaybreak[0]\notag\\
		&\Psi'\left(c_b;z,u^2\right)\triangleq\frac{\partial\Psi\left(c_b;z,u^2\right)}{\partial z}=\frac{\phi\left(\frac{z-r_{b-1}}{u}\right)-\phi\left(\frac{z-r_{b}}{u}\right)}{ u }\notag.
	\end{align}  \hfill\ensuremath{\blacksquare}
\end{Proposition}

\begin{Remark}\label{R2}
	In the OFDM system with infinite-precision quantization, parallel data are transmitted over $N$ mutually orthogonal subchannels.
	The signal-to-noise ratio (SNR) of the $j$-th subchannel is $\frac{p_j |h_j|^2}{\sigma^2}$.
	However, the orthogonality among subchannels in the Q-OFDM system cannot be maintained.
	Proposition \ref{Pro:SE} in conjunction with \eqref{eIII_04} reveals that, in the large-system limit, the input-output relationship of the Q-OFDM system employing Algorithm~\ref{A1} can still be decoupled into a bank of \emph{equivalent} AWGN channels corresponding to $N$ subchannels given by
	\begin{equation} \label{eq:eqAWGN}
		x_{j,B}^{{\rm pri}} = \sqrt{\eta^{t}}\sqrt{p_j}h_js_j + w_{j}
	\end{equation}
	for $j=1,\cdots,N$, where $w_{j} \sim {\cal CN}(0,1)$. We refer to this characteristic as the decoupling principle. The SNR of the equivalent AWGN channel is $p_j |h_j|^2 \eta^{t} $.
	\hfill\ensuremath{\blacksquare}
\end{Remark}

As $B\to\infty$, \eqref{eII_07} is reduced to the OFDM system with infinite-precision quantization.  Let $r_{b-1} = r$ and $r_{b} = r_{b-1} + \rmd r$. As
$B\to\infty$, we obtain $\rmd r \to 0$, which results in $\Phi\left(\frac{z-r_{b-1}}{u}\right)-\Phi\left(\frac{z-r_{b}}{u}\right) \to \frac{\rm d}{\rmd r}
\Phi\left(\frac{z-r}{u}\right)$ and $\phi\left(\frac{z-r_{b-1}}{u}\right)-\phi\left(\frac{z-r_{b}}{u}\right) \to \frac{\rm d}{\rmd r}
\phi\left(\frac{z-r}{u}\right)$. By substituting these relationships into \eqref{eq:SE1} and applying the facts that $\frac{\rm d}{\rmd r}
\Phi\left(\frac{z-r}{u}\right) = \frac{1}{u} \phi\left(\frac{z-r}{u}\right)$ and $\frac{\rm d}{\rmd r} \phi\left(\frac{z-r}{u}\right) =
\left(\frac{z-r}{u^2}\right) \phi\left(\frac{z-r}{u}\right)$, we can obtain
\begin{equation} \label{eq:chi_UnQ}
	\vartheta^{t} = \frac{1}{ \sigma^2 + \nu^{t} }.
\end{equation}
Substituting (\ref{eq:chi_UnQ}) into (\ref{eq:SE2}), we obtain $\eta^{t} = 1/\sigma^2$ for any iteration index $t$. The resulting SNR is perfectly consistent with
that in the infinite-precision OFDM system. Consequently, the parameter $1/\eta^{t}$ can be served as an \emph{equivalent} noise power of the Q-OFDM system, and
$\eta^{t} \leq  1/\sigma^2$.

With the decoupling principle, we can easily predict several fundamental performance metrics, such as MSE, SER, and mutual information, of the Q-OFDM system
without performing time-consuming Monte Carlo simulations. For example, we determine that $\mathrm{mmse}(|h'_j|^2\eta^{t})$ predicts the per-component MSE of
$\mathbf{s}$ at the $t$-th iteration. If the data symbol is drawn from the $M$-QAM constellation, then the SER at the $t$-th iteration can be obtained analytically by \cite{Proakis2007}
\begin{equation} \label{eq:SER_Analysis_QPSK}
		\mathrm{SER}^t=\frac{1}{N}\sum\limits_{j=1}^{N}4\left(1-\frac{1}{\sqrt{M}}\right)Q\left(\sqrt{g_M|h'_j|^2\eta^{t}}\right)\times\left[1-\left(1-\frac{1}{\sqrt{M}}\right)Q\left(\sqrt{g_M|h'_j|^2\eta^{t}}\right)\right],
\end{equation}
where $g_M=\frac{3}{M-1}$.
Clearly, the decoupling principle and the SE equations are useful for performance optimization. For example, the decoupling principle facilitates the allocation of power among $N$ subchannels to optimize some performance metrics, which will be discussed in the subsequent subsection.

\begin{Remark}\label{R4}
	The argument from statistical mechanics (see, e.g., \cite{Shinzato-09JPA,Krzakala-12JSM}) shows that the performance metrics of the Bayesian MMSE estimator, such
	as the MSE of $\mathbf{s}$, correspond to the saddle points of the average free entropy, which is defined as
	\begin{equation} \label{eq:muInfF}
		\mathcal{F} =  -\frac{1}{N}  \mathrm{E}\left[ \log { \mathrm{P}(\mathbf{q} ;\mathbf{h}')} \right],
	\end{equation}
	where the expectation is taken w.r.t. the marginal likelihood in \eqref{eq:MQ}.
	For a review of the statistical mechanics methods applied to high-dimensional inference, please refer to \cite{Advani-13JSM}.
	The calculation of $\mathcal{F}$ and its saddle points are given in Appendix \ref{AP2}.
	The saddle points of $\mathcal{F}$ expressed in \eqref{eq:deftqu}--\eqref{eq:defchix} in Appendix \ref{AP2} are identical to those of the SE equations \eqref{eq:SE1}--\eqref{eq:SE3}, by substituting
	$\frac{1}{\chi_{s}}=\nu$ and $\tilde{q}_{s}=\eta$ into \eqref{eq:deftqu}--\eqref{eq:defchix}.
	This result indicates that Algorithm~\ref{A1} can yield the same estimate as direct integration in \eqref{eq:postMean_x} as the Bayesian MMSE estimator does. \hfill\ensuremath{\blacksquare}
\end{Remark}

\subsection{Power Allocation}

In a frequency-selective fading channel, the channel gains among different subchannels widely vary.
Under the low-precision quantization scenario, data sent from the weaker subchannels tend to be lost because of strong ICI from the stronger subchannels, which leads to a high error floor.
In this subsection, we develop a PA scheme to further improve the SER performance.

Recall from Remark \ref{R2} that the input-output relationship of the Q-OFDM system can be decomposed into a bank of AWGN channels corresponding to $N$ subchannels with SNR $p_j |h_j|^2 \eta$ for $j=1,\cdots,N$.
With this decoupling principle, we can allocate the total power $\sum_{j=1}^{N} p_j = N\bar{P}$ among the $N$ equivalent AWGN channels to optimize some performance metrics.
In particular, we consider the subchannel power allocation that minimizes the SER.
From \cite[Proposition 1]{Dabeer-2013TCOM}, the SER for a $M$-QAM OFDM system under a given channel realization $\{ h_j \}$ and noise power $\eta^{-1}$ is given by
\begin{equation}  \label{eq:SER0}
	\mathrm{SER}=4S-O(S^2),
\end{equation}
where $O(\cdot)$ is the big O notation and
\begin{equation}\label{eq:SER1}
	S={\left(1-\frac{1}{\sqrt{M}}\right)}\frac{1}{N}\sum\limits_{j=1}^{N}Q{\left( \sqrt{g_Mp_j|h_j|^2\eta} \right)}.
\end{equation}

The SER expression is dominated by the first term which is found to be a good approximation \cite{Dabeer-2013TCOM}. Therefore, our goal is to derive the optimal PA
$\{p_j\}_{j=1}^{N}$ that minimizes the dominant term in \eqref{eq:SER1} under the constraint $\sum_{j=1}^{N} p_j = N\bar{P}$. Hereinafter we set $\bar{P}=1$ to simplify the set of simulation parameters. When $\bar{P}=1$, the parameter $\sigma^2$ can be set as the reciprocal of target SNR, and thus normalizing channel gain $g_i$s is easier. However, solving the above problem directly involves an iterative procedure to obtain the solution of $N$ nonlinear equations, which suffers from slow convergence and high computational complexity \cite{Wang-2003GLCOM}.
Thus we resort to an approximation for the Q-function given by $Q(x) \approx \frac{1}{2}\exp\left(-\frac{x^2}{2} \right)$ \cite{Proakis2007}. Accordingly, we formulate the PA problem as
\begin{equation}
	\begin{aligned}
		&\min\limits_{\{p_j\}_{j=1}^{N}\geq0} & & \sum\limits_{j=1}^{N}\exp{\left( -\frac{g_Mp_j|h_j|^2\eta}{2} \right)}, \\
		&{\rm subject~to} && \sum_{j=1}^{N} p_j = N.
	\end{aligned}
\end{equation}
Define the Lagrangian function as
\begin{equation} \label{eq:LagrangianFun}
	\calL=\sum\limits_{j=1}^{N}\exp{\left( -\frac{g_Mp_j|h_j|^2\eta}{2} \right)}+\lambda{\left(\sum_{j=1}^{N} p_j - N\right)},
\end{equation}
where $\lambda$ is the Lagrange multiplier. Equating the partial derivatives of $\calL$ w.r.t. $\{p_j\}_{j=1}^{N}$ to zero, we obtain\footnote{In fact, $\lambda$ in \eqref{eq:PowerAllocation} is not identical to that in \eqref{eq:LagrangianFun}. We hope this slight abuse of notation will cause no
	confusion.}
\begin{equation} \label{eq:PowerAllocation}
	p_j = \left( \frac{{\ln {{\left| {{h_j}} \right|}^2} + \lambda}}{{\gamma {{\left| {{h_j}} \right|}^2}}}\right) ^{+},
\end{equation}
where $(x)^{+}\triangleq\max\{x,0\}$, $\gamma=\frac{g_M\eta}{2}$, and $\lambda$ is the parameter selected to satisfy the constraint $\sum_{j=1}^{N} p_j = N$. This PA is called as the approximate minimum symbol error rate (AMSER) scheme. We develop a process that
resembles water filling to determine $\{p_j\}_{j=1}^{N}$ and $\lambda$, as expressed in \eqref{e_IV1} and \eqref{e_IV2} in Algorithm \ref{A2}. We let $\mathcal{K}$
be the set of subchannel indices with non-zero power with initialization $\{1,2,\cdots,N\}$. For a given $\mathcal{K}$, $\lambda$ can be computed with
\eqref{e_IV11}. If $\min_{j\in\mathcal{K}}\ln|h_j|^2\geq-\lambda$ is satisfied, then the process is terminated. Otherwise, we remove the subchannel
$j_0=\arg\min_{j\in\mathcal{K}}\ln|h_j|^2$ from $\mathcal{K}$ and repeat the process.
\begin{algorithm}[!ht]
	\caption{AMSER Power Allocation\label{A2}}
	\footnotesize
	\KwData{$p_j^0=1$ for $j=1,2,\cdots,N$ and $\nu^0=1$\;}
	\For{$t=1:T_{max}$}{
		\begin{subequations} \label{e161014}
			\begin{align}
			&v^t_x = \frac{1}{N}\sum\limits_{j=1}^{N}p_j^{t-1}|h_j|^2,\label{e161014_01}\\
			&\vartheta^{t}=\frac{1}{2}\sum\limits_{b=1}^{2^B}\int_{-\infty}^{\infty}\frac{\left[\Psi'\left(c_b;\sqrt{\frac{v^t_x -\nu^{t-1}}{2}}z,\frac{\sigma^2+\nu^{t-1}}{2}\right)\right]^2}{\Psi\left(c_b;\sqrt{\frac{v^t_x -\nu^{t-1}}{2}}z,\frac{\sigma^2+\nu^{t-1}}{2}\right)}\mathrm{D}z,\displaybreak[0]\label{e161014_02}\\
			&\eta^{t}=\frac{1}{(\vartheta^{t})^{-1}-\nu^{t-1}},\displaybreak[0]\label{e161014_03}
			\end{align}
		\end{subequations}
		AMSER Power Allocation:\\
		\textbf{Initialization:} $\mathcal{K}=\{1,2,\cdots,N\}$, $\gamma=\frac{g_M\eta^{t}}{2}$\\
		\While{$(1)$}{
			\begin{subequations}\label{e_IV1}
				\begin{align}
				&\lambda {\rm{ = }}\frac{{\gamma-\frac{{\rm{1}}}{N}\sum\nolimits_{j \in \mathcal{K}} {\frac{{\ln {{\left| {{h_j}} \right|}^2}}}{{{{\left| {{h_j}} \right|}^2}}}} }}{{\frac{{\rm{1}}}{N}\sum\nolimits_{j \in \mathcal{K}} {\frac{1}{{{{\left| {{h_j}} \right|}^2}}}} }},\label{e_IV11}\\
				&p_j^{t} = \frac{{\ln {{\left| {{h_j}} \right|}^2} + {\lambda}}}{{\gamma {{\left| {{h_j}} \right|}^2}}},~j\in\mathcal{K}\label{e_IV12}.
				\end{align}
			\end{subequations}
			\eIf{$\min_{j\in\mathcal{K}}\ln|h_j|^2<-\lambda$}{
				\begin{subequations}\label{e_IV2}
					\begin{align}
					&j_0= \mathop{{\rm argmin}}\limits_{j\in\mathcal{K}}\ln|h_j|^2;\\
					&\mathcal{K}=\mathcal{K}\setminus {j0};\\
					&p_{j_0}^{t}=0;
					\end{align}
				\end{subequations}
			}
			{\textbf{break}}
		}
		\begin{equation}\label{e_IV3}
		\nu^{t}=\left(\frac{1}{\frac{1}{N}\sum\limits_{j=1}^{N}p_j^{t}|h_j|^2\mathrm{mmse}(p_j^{t}|h_j|^2\eta^{t})}-\eta^{t}\right)^{-1}.
		\end{equation}
	}
\end{algorithm}

Notably, $\eta$ cannot be directly determined before the PA process because $\eta$ is a function of the allocated power $\{p_j\}_{j=1}^{N}$. Therefore, we embed
AMSER PA process into the iteration of SE equations and obtain Algorithm \ref{A2}. Specifically, in the $t$-th iteration, we compute $\eta^{t}$ through \eqref{e161014_01} to
\eqref{e161014_03} based on the allocated power in the $(t-1)$-th iteration $\{p^{t-1}_j\}_{j=1}^{N}$. Then, with the fixed $\eta^{t}$, we obtain the power
$\{p^{t}_j\}_{j=1}^{N}$ in \eqref{e_IV1} and \eqref{e_IV2} and $\nu^{t}$ in \eqref{e_IV3}. The algorithm requires adapting $\eta^{t}$ and $\{p^{t}_j\}_{j=1}^{N}$
separately in an iterative manner.

As $B\to\infty$, we obtain the parallel channels \eqref{eq:eqAWGN} with $\eta^{t} = 1/\sigma^2$ by the argument following Remark 3. In this case, Algorithm
\ref{A2} is reduced to the AMSER PA performed for $N$ subchannels of the OFDM system with infinite-precision quantization as proposed in \cite{Wang-2003GLCOM}.

\subsection{Computational Complexity}

The computational complexity of the GTurbo-based detector, i.e., Algorithm 1, is dominated by matrix multiplications in \eqref{eA1_06} and \eqref{e0411_08b}.
Fortunately, they can be implemented with fast Fourier transform (FFT) processors with computational complexity $\mathcal{O}(N\log_2N)$. The detector also converges within a few iterations, as discussed later in Section VI. The real bottleneck of the detector implementation comes from the computation of $\Phi(x)=1-Q(x)$, which requires deriving the integral of Gaussian function in \eqref{eA1_01} and \eqref{eA1_02}. The hardware-friendly approximation for $Q(x)$ and the pipelined and folding hardware architecture of a GTurbo algorithm have been proposed in \cite{CZhang2016SiPS}. The simulation results in \cite{CZhang2016SiPS} further demonstrate that the fixed-point setting combined with the Q-function approximation only introduce slight performance degeneration to the original floating-point simulation. This complexity analysis is also valid for the channel estimation algorithm presented in the next section.

Algorithm 2 is for power allocation. The maximum possible number of inner iteration is $N$.
Through extensive simulation, we find that Algorithm 2 typically converges within 10 outer iterations. Therefore, the computational complexity of Algorithm 2 is $\mathcal{O}(N)$.
Furthermore, the integral operation in \eqref{e161014_02} and \eqref{e_IV3} can be generally acquired using a look-up table (LUT).
Consequently, the two algorithms are computational efficiently and hardware friendly.

\section{Channel Estimation}

\begin{figure}[!htb]
    \centering
    \includegraphics[scale=0.35]{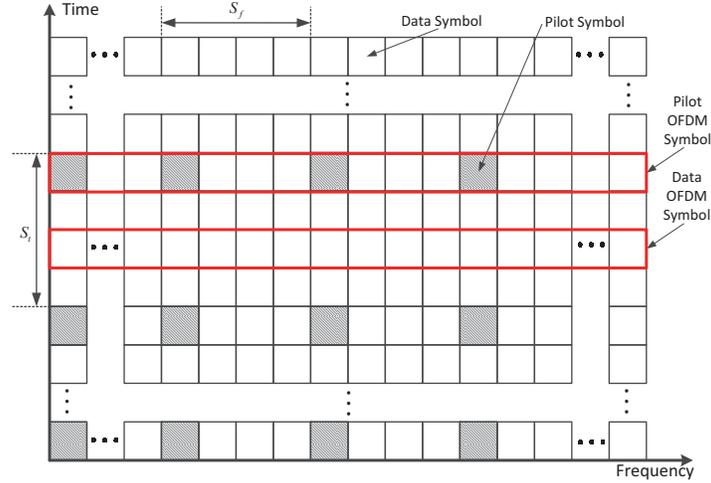}
    \caption{Comb-type pilot arrangment\label{F2}.}
\end{figure}

In this section, we develop a pilot-based channel estimation approach to obtain the CSI based on the GTurbo framework in Algorithm~\ref{A1}. The pilot sequences are known at the
transmitter and receiver sides. In this study, we employ the comb-type pilot arrangement, as shown in Fig.~\ref{F2}, in which the pilot signals are uniformly
inserted into the subchannels of an OFDM symbol. 

We denote the interval of adjacent subchannels containing pilot signals by $S_f$. We use
$\mathcal{X}=\{1,2,\cdots,N\}$ to denote the index set of all subchannels, and we use $\mathcal{X}_p\subseteq\mathcal{X}$ and $\mathcal{X}_d\subseteq\mathcal{X}$
to denote the index subset of the subchannels containing pilot and data symbols, respectively. The pilots are transmitted periodically every $S_t$ OFDM symbols.
During each interval of $S_t$ OFDM symbols, only one OFDM symbol contains the pilot signals (called the pilot OFDM symbol), whereas the other $S_t-1$ OFDM symbols are dedicated to
data transmission. Notably, the pilots are contaminated by the data subchannels because of the use of the coarse quantization which results in severe ICI. The
conventional pilot-based channel estimation schemes for OFDM systems do not consider this effect and thus cannot work well.

Algorithm~\ref{A3} is designed only for pilot OFDM symbols to output the estimated channel $\hat{\mathbf{h}}$ and data $\hat{\mathbf{s}}$. The estimated channel
in the pilot OFDM symbol is subsequently utilized as the CSI for data detection by applying Algorithm~\ref{A1} to the remainder of the OFDM symbols dedicated to
data transmission in each interval of $S_t$ OFDM symbols. Moreover, the estimated channel is sent back to the transmitter for PA (i.e., Algorithm~\ref{A2}). Notably the power can be equally allocated in the pilot OFDM symbol.

\begin{figure}[!h]
	\centering
	\includegraphics[scale=0.75]{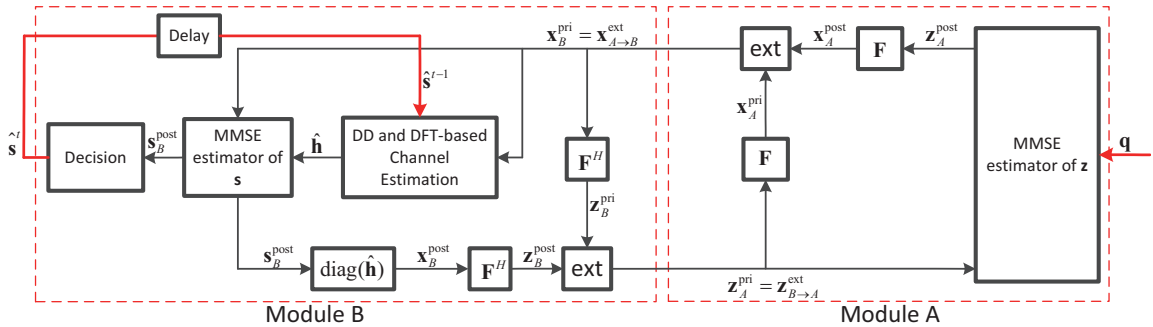}
	\caption{The GTurbo-based channel estimation and the data detection algorithm. The ``ext'' block represents the extrinsic information computation. The block of a certain matrix represents the left-multiplying the input vector by the matrix in the block. \label{F3}}
\end{figure}

The block diagram of Algorithm~\ref{A3} is illustrated in Fig. \ref{F3}. The operations of Module A in Algorithm~\ref{A3} is identical to that in
Algorithm~\ref{A1}. The output of Module A $\mathbf{x}_B^{{\rm pri}}$ can be viewed as the equivalent channel in \eqref{eIII_04}, where each subchannel component
is expressed by the product of the transmitted signal and the channel frequency response at the corresponding subchannel plus an AWGN with power $v_B^{\rm pri}$. This
decoupling property facilitates the subsequent channel estimation and data detection. In particular, through $\mathbf{x}_B^{{\rm pri}}$, we can process the pilot
subchannel $\mathcal{X}_p$ and the data subchannel $\mathcal{X}_d$ separately. For example, \eqref{eA3_02} employs the least squares method to obtain an initial
channel estimation.
\begin{algorithm}[!ht]
	\caption{GTurbo Channel Estimation and Data Detection\label{A3}}
	\footnotesize
	\KwData{$\mathbf{z}_A^{{\rm pri}}=\mathbf{0}_{N\times 1}$, $v_A^{{\rm pri}}=1$\;}
	\For{$t=1:T_{max}$}{
		\textbf{Module A:}\\
		Identical to \eqref{eA1_01}--\eqref{eA1_02}, \eqref{eA1_04}--\eqref{eA1_06}.\\
		\textbf{Module B:}\\
		(3) Coarse channel estimation:\\
		\eIf{$t=1$}{\begin{equation}\label{eA3_02}
			{{\tilde h}_j} = \left\{ {\begin{array}{*{20}{l}}
				{S_f\frac{x_{j,B}^{\rm pri}}{s_j},}&{j \in {\mathcal{X}_p},}\\
				{0,}&{j \in {\mathcal{X}_d},}
				\end{array}} \right.
			\end{equation}}
		{\begin{equation}\label{eA3_03}
			{{\tilde h}_j} = \frac{x_{j,B}^{\rm pri}}{\hat{s}^{t-1}_j},~~~{j \in {\mathcal{X}},}
			\end{equation}}
		
		(4) Refinement of channel estimation:
		\begin{subequations} \label{eA3}
			\begin{align}
			&\tilde{\mathbf{g}}=\mathbf{F}^H\tilde{\mathbf{h}},\label{eA3_04}\\
			&{{\hat g}_i} = \left\{ {\begin{array}{*{20}{l}}
				{{{\tilde g}_i},}&{i \le L,}\\
				{0,}&{{\rm{otherwise,}}}
				\end{array}} \right.\label{eA3_05}\\
			&\hat{\mathbf{h}}=\mathbf{F}\mathbf{\hat{g}}.\label{eA3_06}
			\end{align}
		\end{subequations}
		(5) Data detection:
		\begin{subequations}
			\begin{align}
			&s_{j,B}^{\rm post}=\mathrm{E}\left\{s_{j}\mid \hat{h}_{j}, x_{j,B}^{{\rm pri}}\right\},~{j \in {\mathcal{X}},} \label{e0530_06a}\\
			&v_{j,B}^{\rm post}=\mathrm{var}\left\{s_{j}\mid \hat{h}_{j}, x_{j,B}^{{\rm pri}}\right\},~{j \in {\mathcal{X}},} \label{e0530_06b}\\
			&{{\hat s}^t_j} = \left\{ {\begin{array}{*{20}{l}}
				{s_j,}&{j \in {\mathcal{X}_p},}\\
				{\min\limits_{s\in\mathcal{S}} \left|s-s_{j,B}^{\rm post}\right|^2,}&{j \in {\mathcal{X}_d},}
				\end{array}} \right.\label{eA3_06c}
			\end{align}
		\end{subequations}
		(6) Compute the extrinsic mean/variance of $\mathbf{z}$:
		\begin{subequations}
			\begin{align}
			&x_{j,B}^{\rm post} = \hat{h}_{j}s_{j,B}^{\rm post},\label{e0531_07a}\\
			&v_{B}^{\rm post}=\frac{1}{N}\sum\limits_{j=1}^{N}|{\hat h}_j|^2v_{j,B}^{\rm post}\label{e0531_07b},\\
			&v_A^{{\rm pri}}=v_{B}^{\rm ext}=\left(\frac{1}{v_{B}^{\rm post}}-\frac{1}{v_B^{{\rm pri}}}\right)^{-1},\\
			&\mathbf{z}_A^{{\rm pri}}=\mathbf{z}_{B}^{\rm ext}=v_{B}^{\rm ext}\left(\frac{\mathbf{F}^H\mathbf{x}_B^{\rm post}}{v_{B}^{\rm post}}-\frac{\mathbf{F}^H\mathbf{x}_B^{{\rm pri}}}{v_B^{{\rm pri}}}\right).
			\end{align}
		\end{subequations}
	}
\end{algorithm}

Once the initial channel estimate in the first iteration is obtained, the estimated channel is updated using the decision-direct (DD) technique in the subsequent
iterations. Specifically, in the $t$-th iteration, the DD technique uses the detected signal in the $(t-1)$-th iteration $\hat{\mathbf{s}}^{t-1}$ to estimate
$\mathbf{h}$ coarsely in \eqref{eA3_03}. Afterward, we transform the coarsely estimated frequency channel response $\tilde{\mathbf{h}}$ to the time domain in
\eqref{eA3_04} and refine the estimate by eliminating the effect of noise outside the maximum channel delay $L$ in \eqref{eA3_05}. Finally, we transform
$\mathbf{\hat{g}}$ back to the frequency domain in \eqref{eA3_06}.

Subsequently, we use the estimated channel $\hat{\mathbf{h}}$ for data detection. The posteriori mean and variance of the data symbols can be calculated similar
to \eqref{eIII_05} and \eqref{eIII_06} while replacing the exact channel response $h_j$ with estimated channel response $\hat{h}_j$ as shown in \eqref{e0530_06a}
and \eqref{e0530_06b}. For $j\in{\mathcal{X}_d}$, the decision ${{\hat s}^t_j}$ is made according to the rule \eqref{eq:asy_ML}, while for $j\in{\mathcal{X}_p}$, ${{\hat s}^t_j}$ takes the
pilot signal. In step (6) of Algorithm~\ref{A3}, the extrinsic mean and variance of $\mathbf{z}$ are computed and used as the input of Module A. Similar to Algorithm~\ref{A1}, two modules are executed iteratively until convergence.

\section{Simulation Results}

Computer simulations are conducted to evaluate the performance of the proposed algorithms and verify the accuracy of our analysis. In the simulations, the number
of OFDM subchannels is $N=512$ and the number of channel taps is $L=4$. The channel impulse response $g_i$ for $i=1,\cdots,L$ is assumed to be i.i.d. with PDF
$\mathcal{CN}(g_i;0,N/L)$. Each entry of the transmitted symbols $\mathbf{s}$ is drawn from the equiprobable QPSK constellation without specific indication. We set ${\rm E}[|s_j|^2] =
1$ for $j=1,\cdots,N$, thus the average SNR can be given by $1/\sigma^2$. The SER, which is averaged over all subchannels, is obtained through the Monte-Carlo simulations of 1,000 independent channel realizations.

\begin{figure}[!h]\centering	
\subfloat[N=512 under ESPA and AMSER PA]{\includegraphics[scale=0.4]{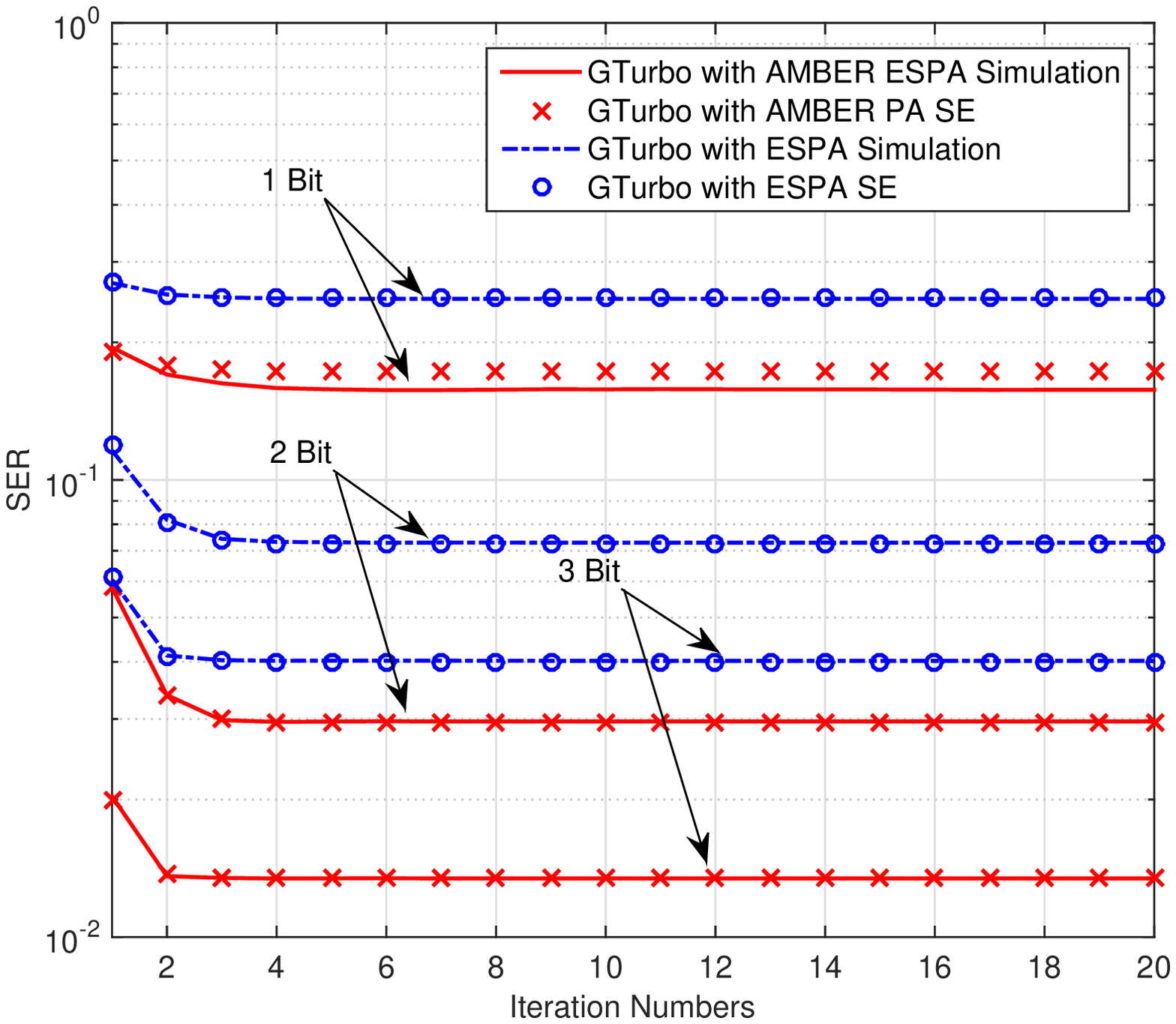}\label{F4a}}
	\subfloat[N=16 and N=32 under ESPA]{\includegraphics[scale=0.4]{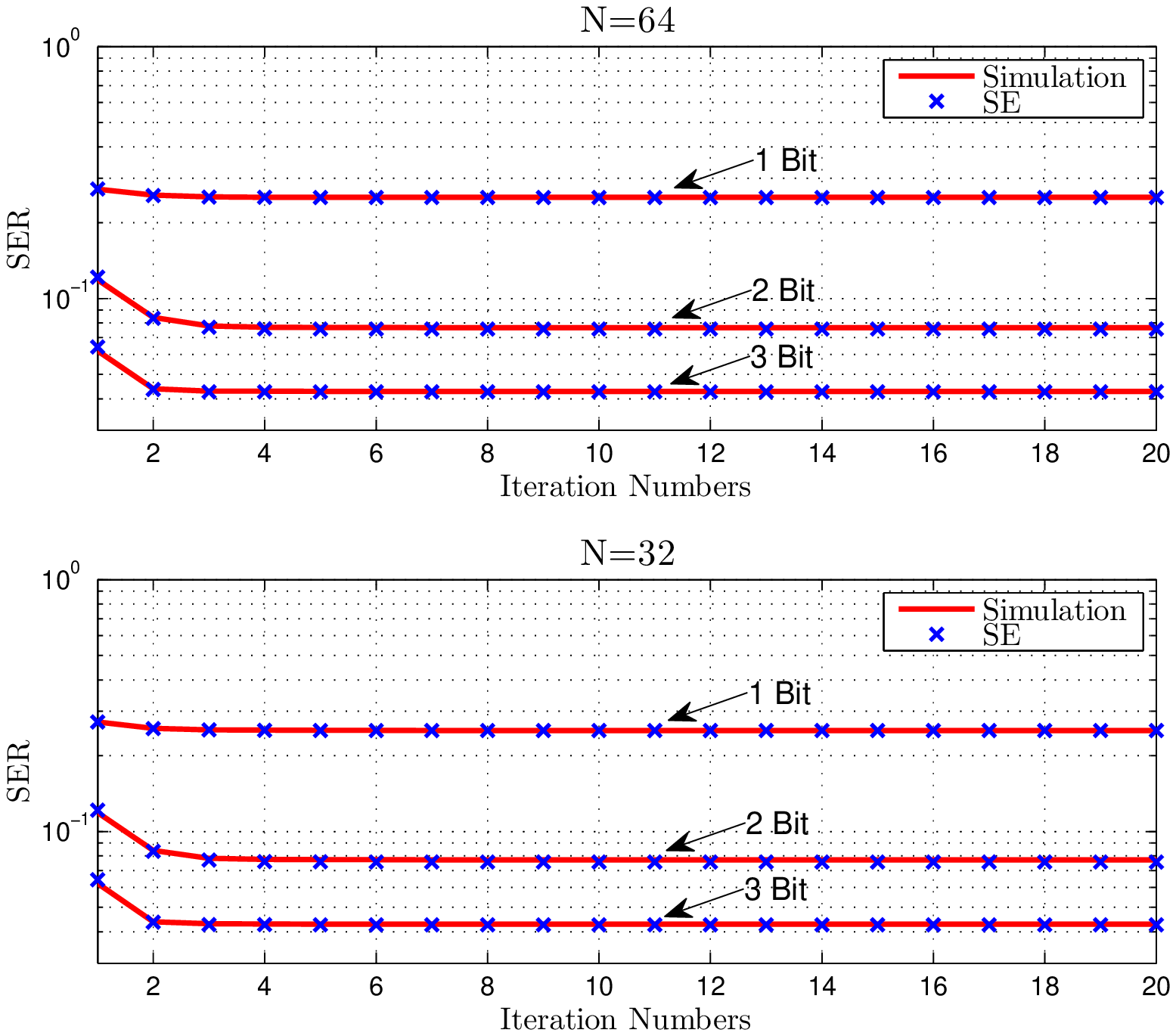}\label{F4b}}
	\caption{SERs versus algorithm iteration of the proposed GTurbo-based detector (i.e., Algorithm~\ref{A1}) under different quantization levels when ${\rm SNR}=15$dB for different subcarrier number $N$.
\label{F6}}
\end{figure}

Fig.~\ref{F4a} shows the SERs versus the iteration numbers of the proposed detector, that is, Algorithm~\ref{A1}, under the quantization precision of 1--3 bits.
The simulated SERs are obtained by the Monte-Carlo simulations of Algorithm 1, while the SE predictions are evaluated using
\eqref{eq:SE_parameters} and \eqref{eq:SER_Analysis_QPSK}.
The SERs under two different PA schemes, i.e., the ESPA and the AMSER PA proposed in Algorithm~\ref{A2}, are evaluated.
Fig.~\ref{F4a} shows that the proposed detector evidently converges within five iterations, and the SE predictions match well with the simulated
results for all quantization settings and PA schemes.
Furthermore, we observe significant SER gaps between the AMSER PA and the ESPA, which validate the effectiveness of the PA scheme proposed in Algorithm~\ref{A2}.
To analyze the asymptotic behavior, we show the simulated and SE results for Algorithm~\ref{A1} with $N=64$ and 32 under ESPA in Fig.~\ref{F4b}. It is shown that the performance of proposed detector is very close to the Bayesian optimal performance in the large system limit, where $N\to\infty$, even for a small number of subcarriers.

\begin{figure}[!t]\centering \label{F5}
	\subfloat[AMSER PA, QPSK]{\includegraphics[scale=0.32]{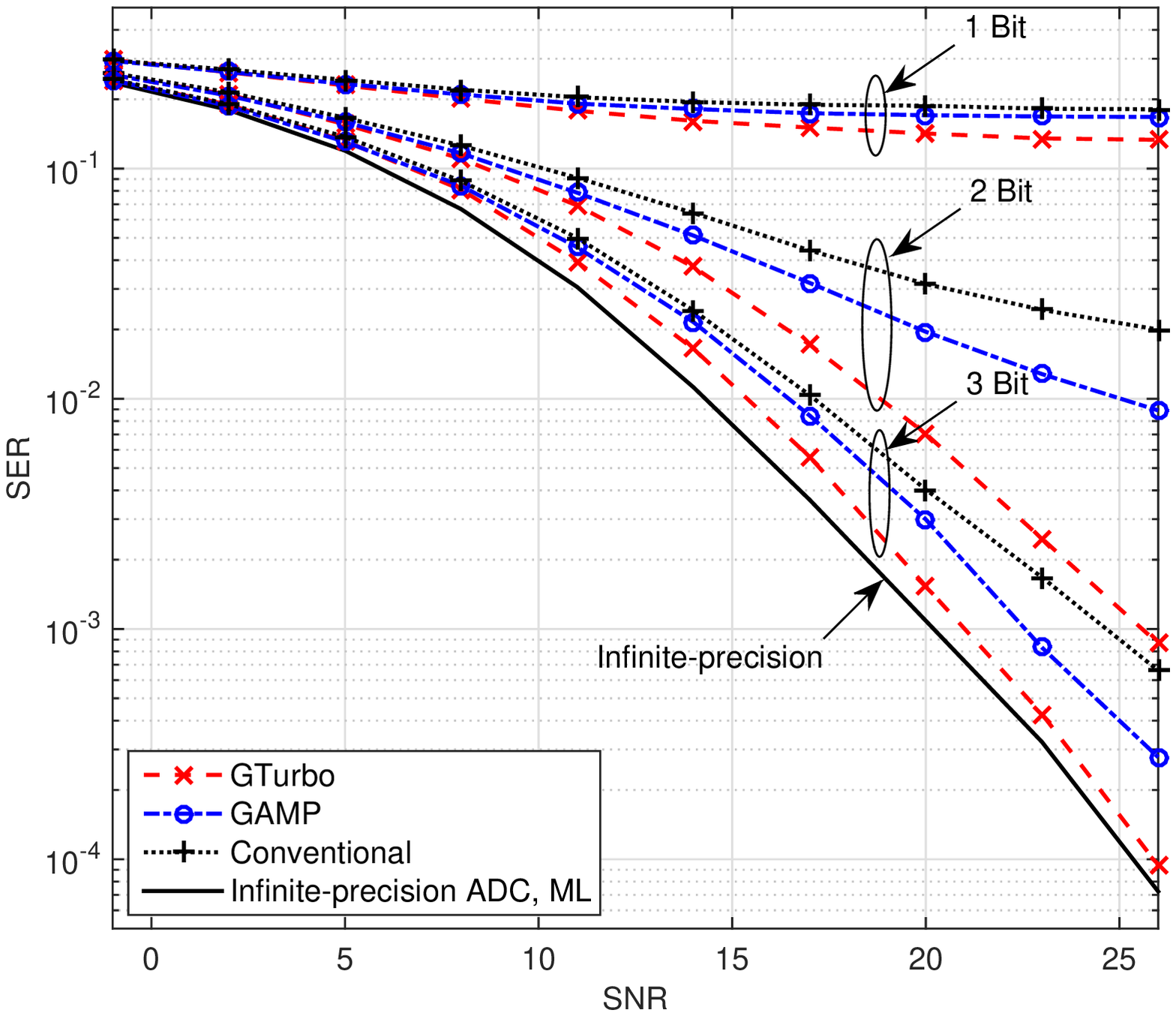}\label{F5A}}
	\hspace{-19pt}
	\subfloat[ESPA, QPSK]{\includegraphics[scale=0.32]{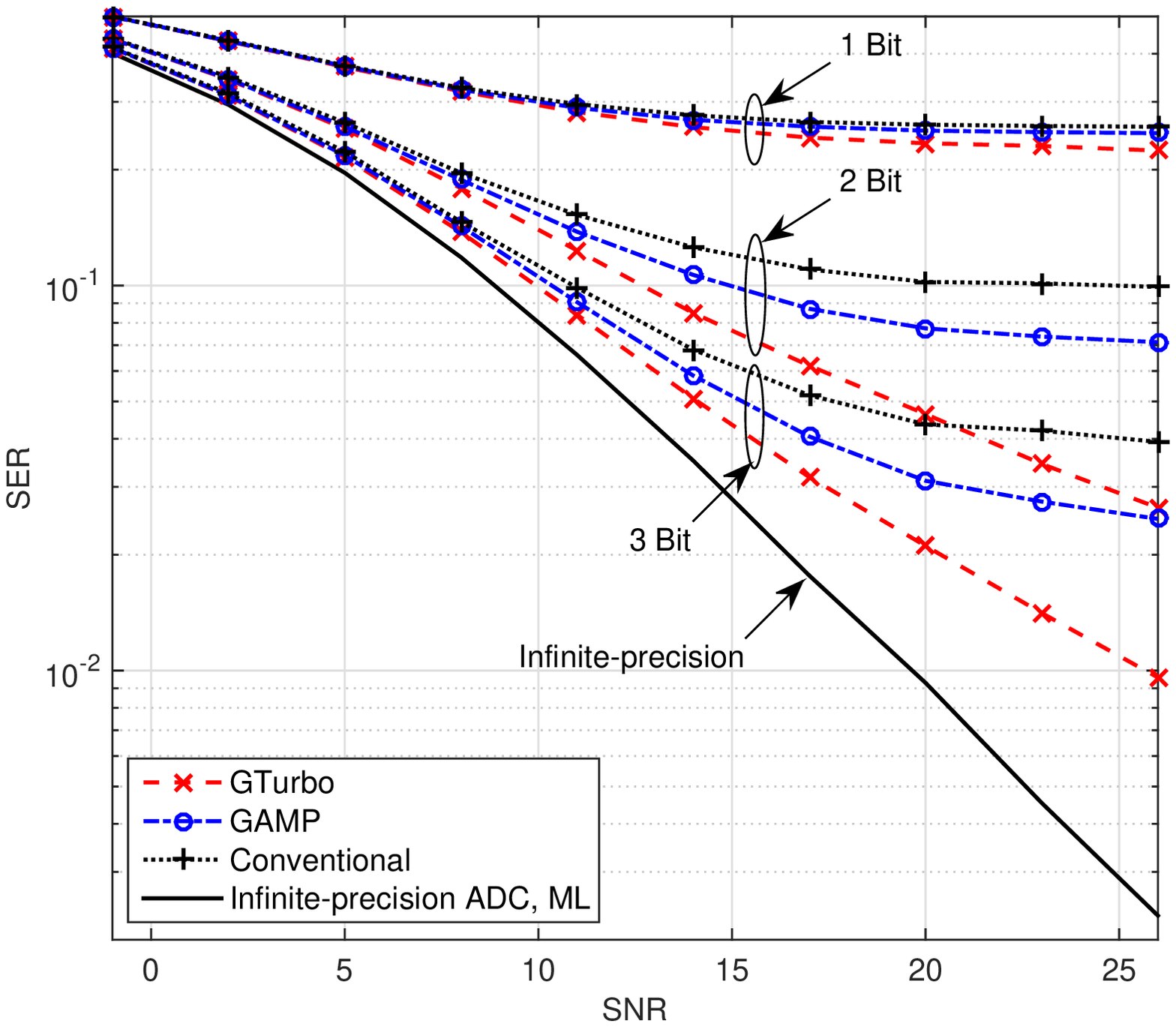}\label{F5B}}
	\hspace{-19pt}
	\subfloat[AMSER PA, 16QAM]{\includegraphics[scale=0.32]{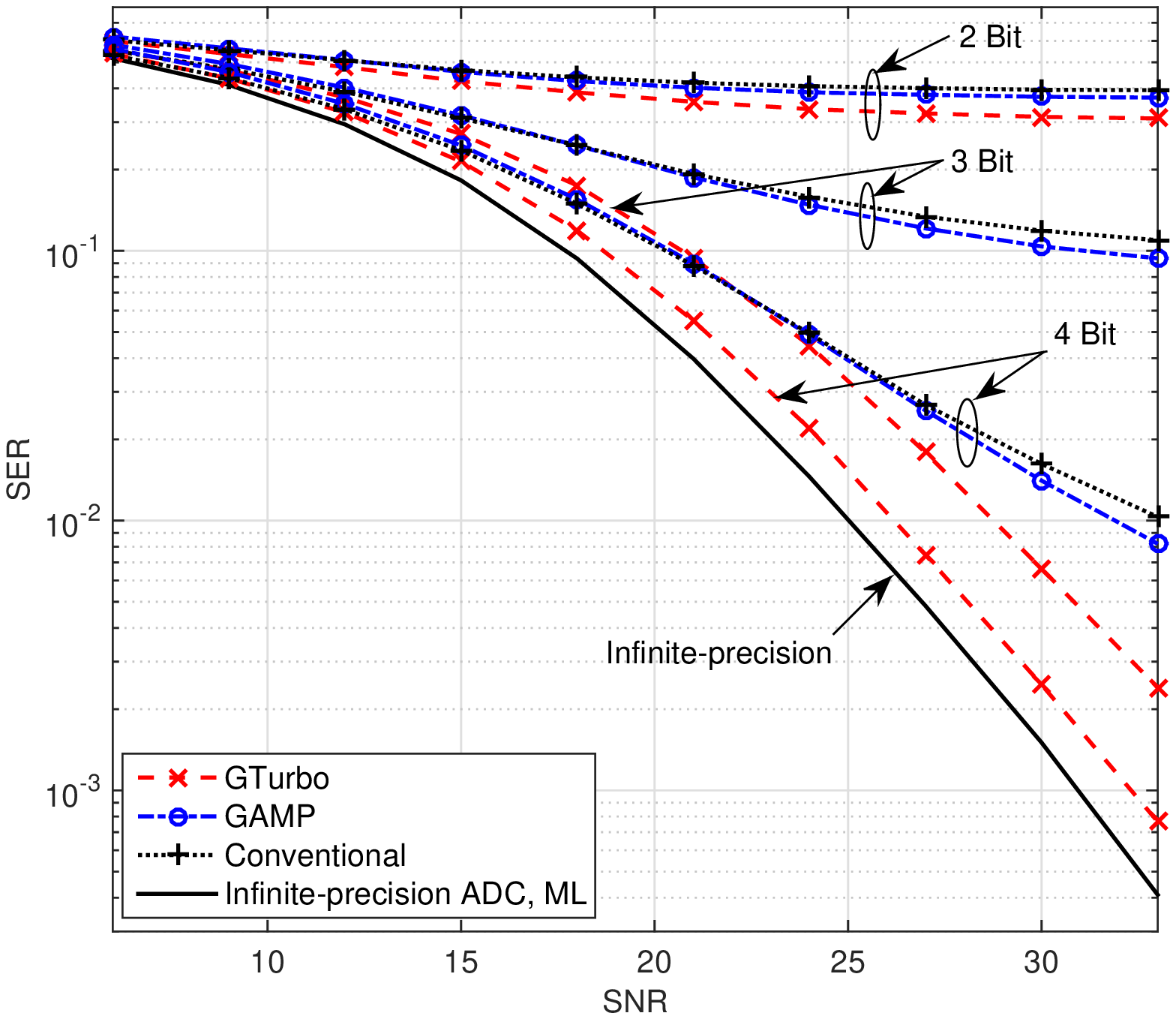}\label{F5C}}
	\caption{SER performance comparisons of the proposed GTurbo-based detector, the GAMP-based detector, and the conventional detector under the perfect CSIR and two different PA schemes and two modulation scheme: a) the AMSER PA scheme for QPSK, and b) the ESPA scheme for QPSK, c) the AMSER PA scheme for 16QAM. \label{F5}}
\end{figure}

Fig. \ref{F5} compares the SERs of the proposed GTurbo-based detector with the existing detectors including the GAMP-based detector \cite{Rangan-2012Arxiv} and the conventional
detector using the one-tap equalizer expressed in (\ref{eq:convML}). The corresponding SERs under the AMSER PA and the ESPA are shown in Figs. \ref{F5A} and
\ref{F5B}, respectively. Notably, the proposed detector significantly outperforms the other two detectors in terms of SER performance. The poor performance
obtained by the conventional detector and the GAMP-based detector can be understood as follows: The conventional detector completely ignores the ICI effect caused
by low resolution ADCs. Although the GAMP-based detector considers the ICI effect, this detector regards the linear transformation matrix of the detection problem \eqref{eII_07} as the i.i.d. entries, and it does not exploit the orthogonality property of the OFDM waveform. Notably, the proposed detector has already achieved the best performance of the Bayesian optimal detector, which
indicates that no further improvement is required. The figures show the optimal SER performance of the OFDM system with infinite-resolution ADCs as the benchmark.
We observe that the SER performance of the GTurbo-based detector with AMSER PA is similar to the optimal performance of the infinite-precision OFDM system.
This result illustrates the feasibility of using very-low-resolution ADCs at the receiver in OFDM systems. Note that
only the sign of real and imaginary parts of the analog received signal the quantized is preserved under 1-bit quantization. The amplitude information of the analog received signal is completely lost. Under such cases with serious non-linear distortion, neither GTurbo- nor GAMP-based detector yields good detection performance without array gain arising from the large-scale antenna array at the receiver as in \cite{Choi-16TCOM}, or involving channel coding.

Particularly, the proposed GTurbo-based detector also works well for high-order modulations such as 16QAM shown in Fig. \ref{F5C}.
When advanced coding techniques, such as \cite{Krzakala-2016Arxiv}, are involved, the transmission of high-order modulation under lower quantization bits and SNR region can be properly supported.
In order to avoid that the key advantages of the proposed detector be obfuscated by other coding technique, we leave this high-order modulation supporting transmission strategies for the future work.
\begin{figure}[!t]\centering
	\subfloat[AMSER PA]{\includegraphics[scale=0.4]{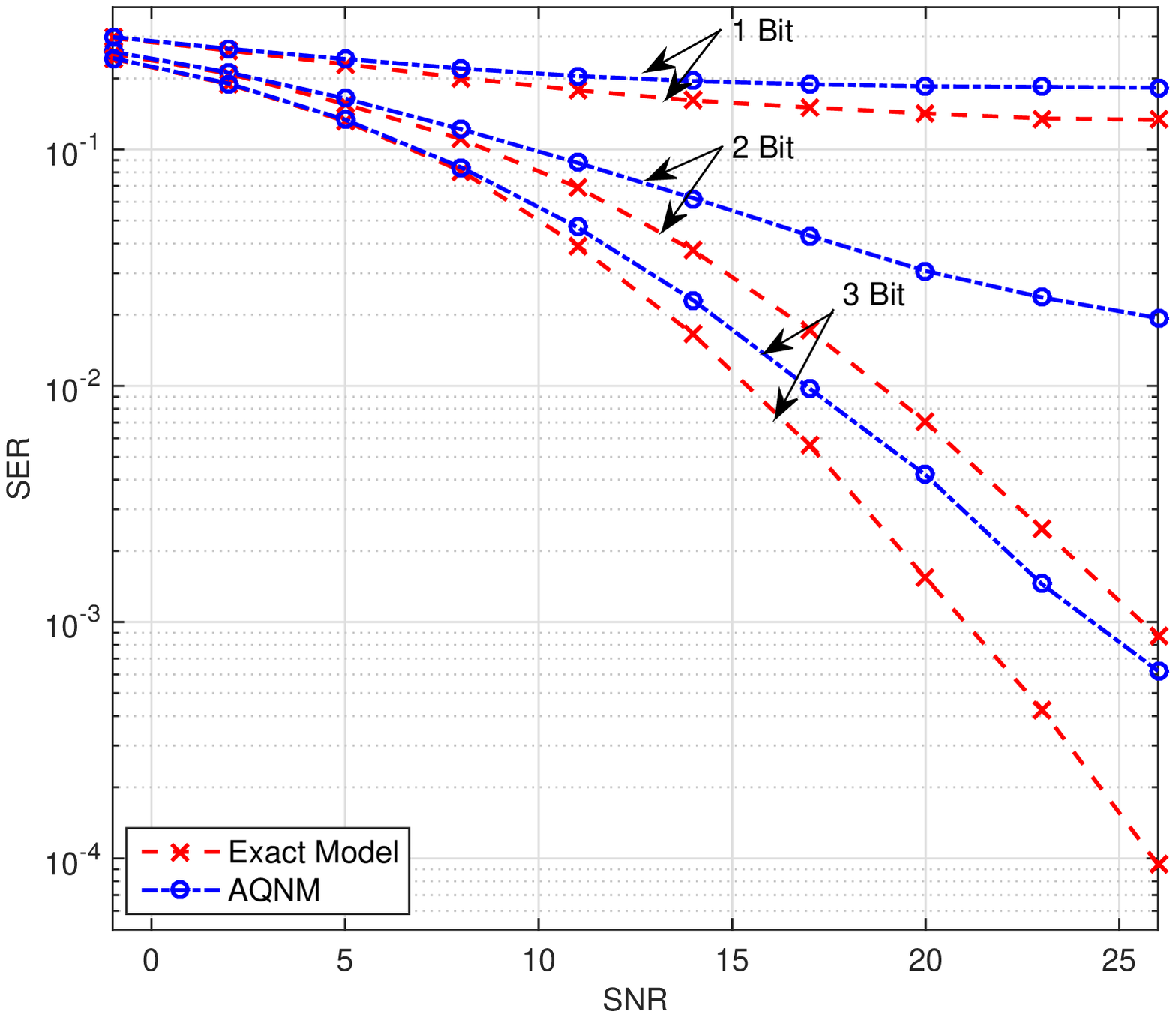}\label{F6A}}
	\subfloat[ESPA]{\includegraphics[scale=0.4]{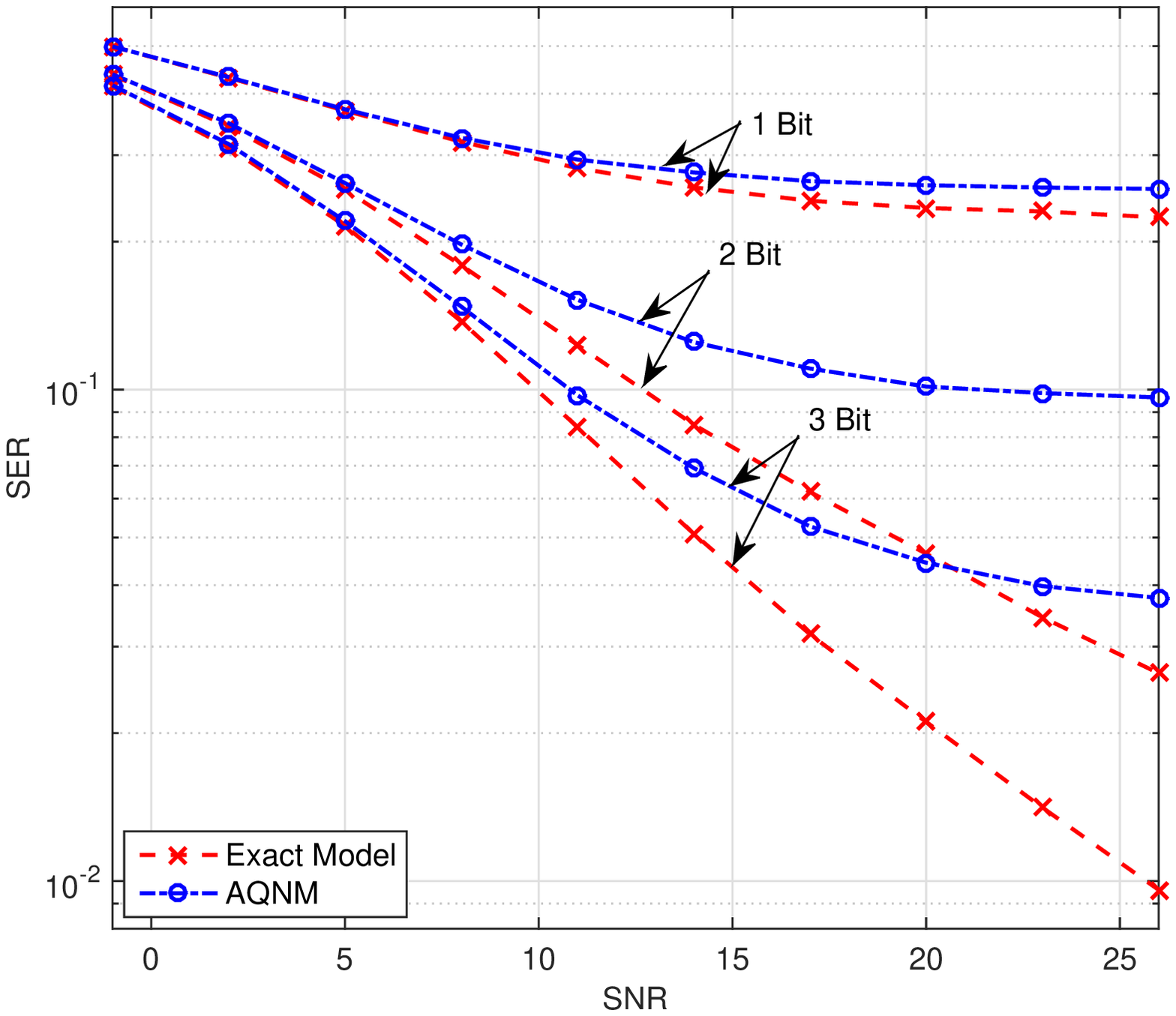}\label{F6B}}
	\caption{SER performance comparisons of  the optimal detector for the exact quantization model and the AQNM under the perfect CSIR and two different PA schemes. \label{F6}}
\end{figure}

In Module A of the GTurbo-based detector, we reconstruct $\mathbf{z}$ from the quantized observation $\mathbf{q}$ using the Bayesian MMSE estimate in \eqref{eA1}.
Another widely used way to deal with quantization noise is to model it as an additive and independent noise, that is, AQNM \cite{SRangan2015ITA}, which allows the use of linear detectors.
Figs. \ref{F6A} and \ref{F6B} compare the optimal detection performances based on the exact quantization model and the AQNM.
Notably, the optimal detection algorithm developed for the AQNM suffers from significant performance loss and severe error floor compared with that for the exact model.
The main reason is that AQNM assumes that the input of the quantizer $y_j$ is a Gaussian variable and approximates the correlated quantization noise by an independent Gaussian noise, which cannot provide a satisfactory approximation to the strongly nonlinear relation  \eqref{eII_07} under the quantization resolution of 1--3 bits.
Furthermore, the comparison of Figs.~\ref{F5A} and \ref{F5B} and that of Figs.~\ref{F6A} and \ref{F6B} illustrate that the use of AMSER PA substantially improves the SER performance.
The decline of SER versus SNR becomes steeper when the PA is performed.
\begin{figure}[!h]\centering
	\subfloat[MSE performance]{\includegraphics[scale=0.4]{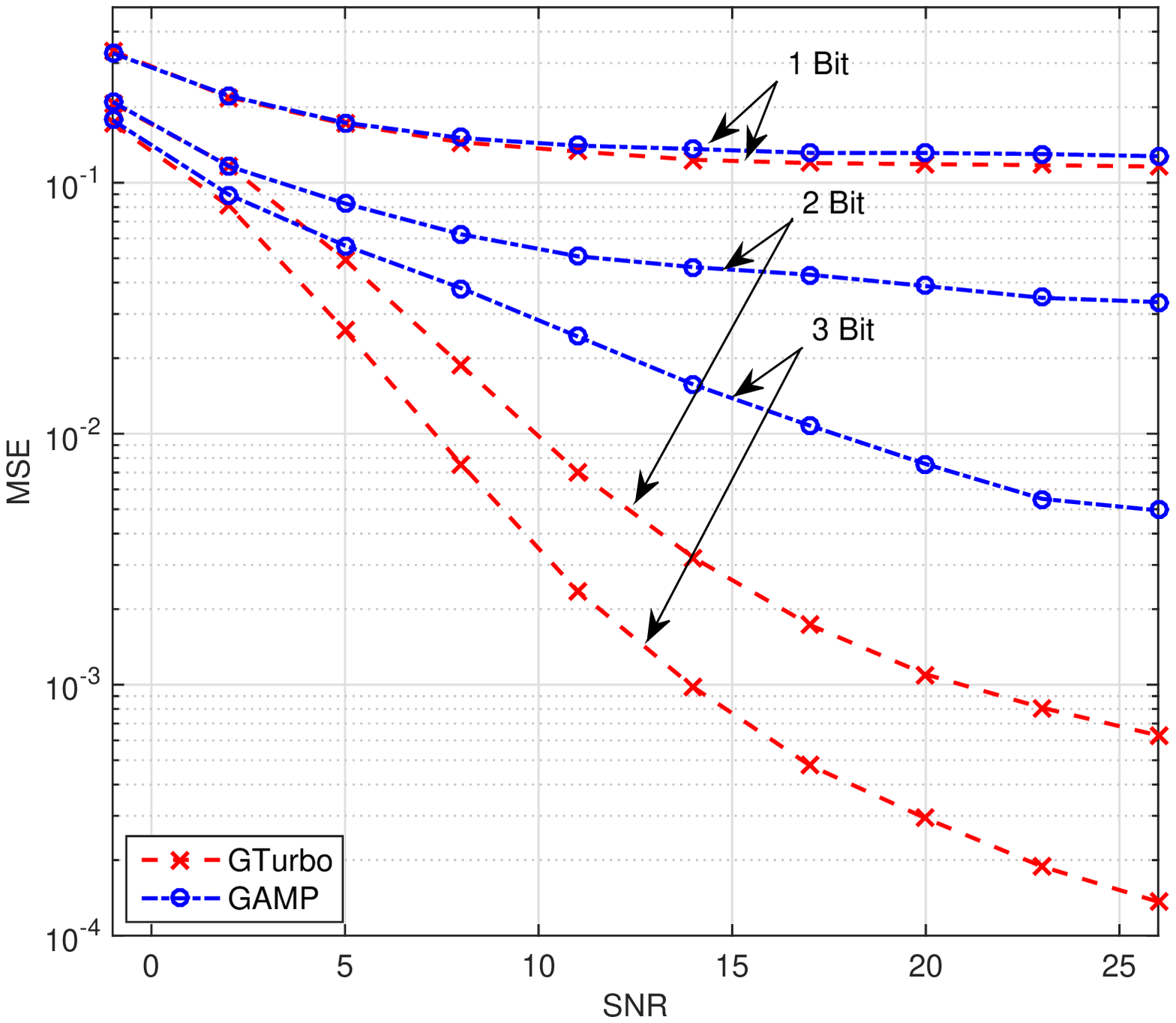}\label{F7A}}
    \vspace{30pt}
	\subfloat[Influence of estimated CSI]{\includegraphics[scale=0.4]{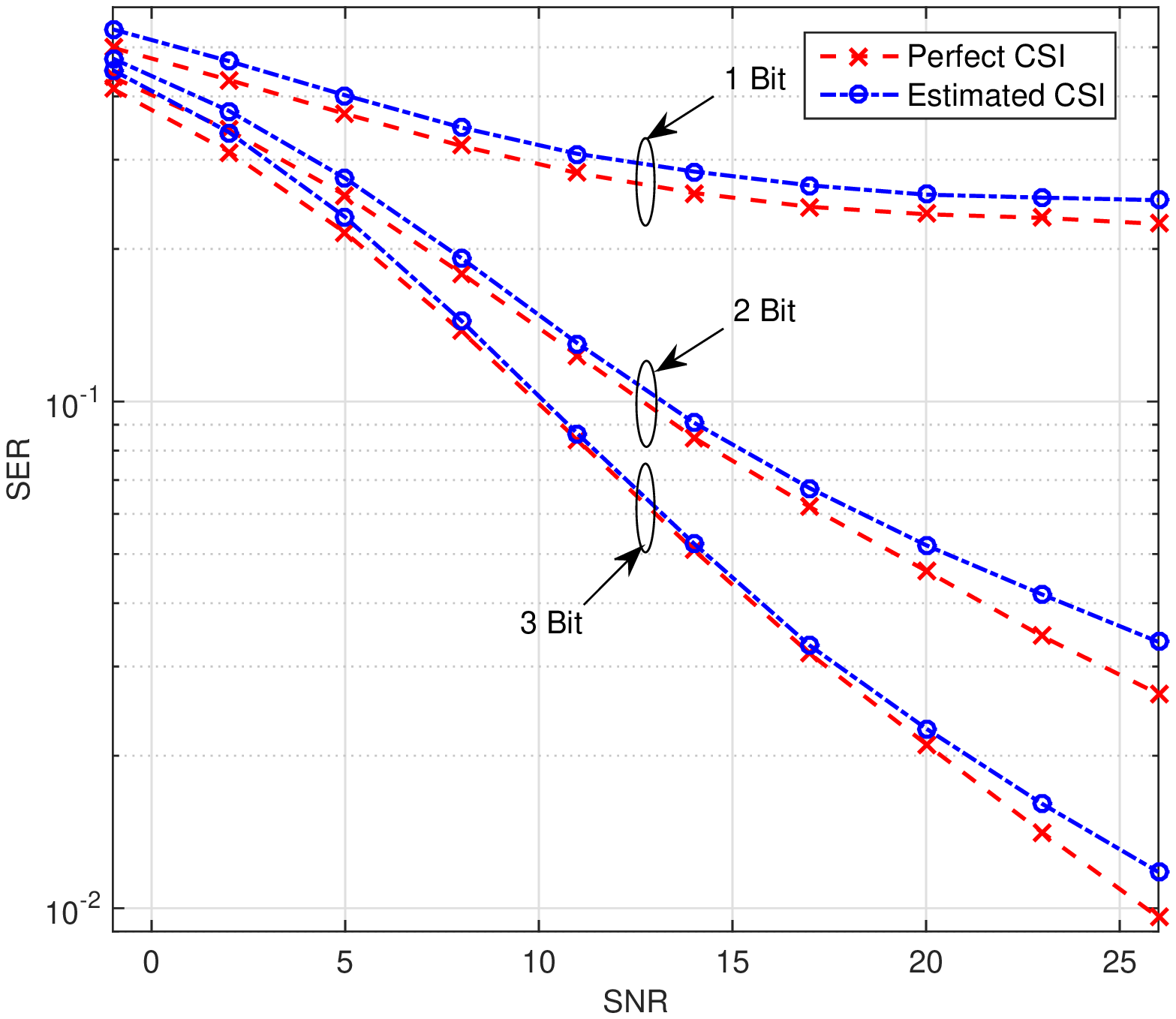}\label{F7b}}
	\vspace{-30pt}
	\caption{Performance evaluations of the channel estimation algorithm: (a) MSE performance of Algorithm~\ref{A3} and GAMP-based channel estimation; (b) SER performance of Algorithm 1 under perfect CSI and estimated CSI.}
\end{figure}

Finally, we examine the channel estimation of the pilot-based OFDM system where the pilot OFDM symbol is arranged as that in Fig. \ref{F2} with $S_f=16$.
The MSE of the channel estimate is defined as $\mathrm{MSE}=\frac{1}{N}\mathrm{E}\left[||\mathbf{h}-\hat{\mathbf{h}}||^2\right]$. Fig. \ref{F7A} shows the MSE of the channel estimation implemented in Algorithm \ref{A3} and the GAMP-based data detection combined with the least square channel estimation method and the refinement technique in \eqref{eA3}.
We observe that the proposed channel estimation significantly outperforms the GAMP-based scheme, particularly for the quantization precision of 2--3 bits.
To further evaluate the performance of the proposed channel estimation algorithm, we compare the detection performance under perfect and estimated CSI, as shown Fig. 7b.
The gap between two cases is comparatively small, especially for 3-bit quantization.
These results justify the feasibility of obtaining high-quality CSI with low-precision ADCs at the receiver without significant pilot overhead.

\section{Conclusion}

We proposed an efficient algorithm for optimal data detection in the Q-OFDM system emerging from mmWave communications.
The SE equations of the proposed detector were derived and shown to be identical to those obtained from the Bayesian optimal detector via the replica theory.
We described the decoupling principle, from which a PA scheme was developed to further improve the SER performance.
Under a unified framework, we also developed a feasible method for channel estimation so that the Q-OFDM detector can be applied to a
practical scenario without perfect CSI.
The simulation results provided the following useful observations:
\begin{itemize}
	\item The algorithm converges rapidly, and its SE prediction is consistent with the simulated result,
	which ensures the quick and efficient performance analysis for the Q-OFDM system.
	\item The proposed PA scheme improves the SER performance significantly and alleviate the error floor compared with the ESPA scheme.
	\item The optimal detector for the Q-OFDM system entails acceptable performance loss compared with that for the infinite-precision case, which confirms the feasibility of the proposed Q-OFDM receiver.
	\item Approximating the input-output relationship of a coarse quantizer by AQNM yields worse detection performance in the Q-OFDM system.
	\item High-quality CSI is available under the Q-OFDM system without significant pilot overhead.
\end{itemize}

\appendices
\section{Proof of Proposition 1}\label{AP1}
In this Appendix, we present the derivation of the SE equations for Algorithm 1 by following \cite{CKWen2016ISIT}. In the large-system  limit where $N\to\infty$,
$v_{A}^{\mathrm{post}}$ in \eqref{eA1_04} converges to the expectation of $v_{j,A}^{\mathrm{post}}$ w.r.t. $z_{j,A}^{\mathrm{pri}}$ and $q_j$ according to the
large-number theorem. For the ease of computation, we first derive the expectation of real part of $\mathrm{var}\left[z^R_j\mid q^R_j\right]$ and add the
expectations of $\mathrm{var}\left[z^R_j\mid q^R_j\right]$ and $\mathrm{var}\left[z^I_j\mid q^I_j\right]$ together. To obtain these expectations, we need the joint
distribution $\mathrm{P}(\mathrm{z_{j,A}^{\mathrm{pri},\mathrm{R}}},q^{\mathrm{R}}_j)$, which can be computed by
$\mathrm{P}(z_{j,A}^{\mathrm{pri},\mathrm{R}},q^{\mathrm{R}}_j)=\int
\mathrm{P}(q^{\mathrm{R}}_j|z_{j,A}^{\mathrm{pri},\mathrm{R}},z^{\mathrm{R}}_j)\mathrm{P}(z_{j,A}^{\mathrm{pri},\mathrm{R}},z^{\mathrm{R}}_j)\mathrm{d}z^{\mathrm{R}}_j$.
The joint distribution of $z_{j,A}^{\mathrm{pri},\mathrm{R}}$ and $z^{\mathrm{R}}_j$ is given by \cite{CKWen2016ISIT}
\begin{equation}\label{e0515_02}
 \mathrm{P}(z_{j,A}^{\mathrm{pri},\mathrm{R}},z^{\mathrm{R}}_j) = \mathcal{N}{\left(z^{\mathrm{R}}_j;z_{j,A}^{\mathrm{pri},\mathrm{R}},\frac{v_{A}^{\mathrm{pri}}}{2}\right)}
 \mathcal{N}{\left(z_{j,A}^{\mathrm{pri},\mathrm{R}};0,\frac{v_x-v_{A}^{\mathrm{pri}}}{2}\right)},
\end{equation}
where $v_x=E(|x_j|^2)=\frac{1}{N}\sum_{j=1}^{N}|h'_j|^2$. Given that $q^{\mathrm{R}}_j$ is independent of $z_{j,A}^{\mathrm{pri},\mathrm{R}}$, we have
$\mathrm{P}(q^{\mathrm{R}}_j|z_{j,A}^{\mathrm{pri},\mathrm{R}},z^{\mathrm{R}}_j)=\mathrm{P}(q^{\mathrm{R}}_j|z^{\mathrm{R}}_j)$; we therefore have the following:
\begin{equation}\label{e0515_03}
\mathrm{P}(q^{\mathrm{R}}_j|z_{j,A}^{\mathrm{pri},\mathrm{R}},z^{\mathrm{R}}_j)=\int_{l(q^{\mathrm{R}}_j)}^{u(q^{\mathrm{R}}_j)}\mathcal{N}{\left(y^{\mathrm{R}}_j;z^{\mathrm{R}}_j,\frac{\sigma^2}{2}\right)}\mathrm{d}y^{\mathrm{R}}_j.
\end{equation}

Combining \eqref{e0515_02} and \eqref{e0515_03}, we have the following:
\begin{equation}\label{key}
\begin{aligned}
\mathrm{P}&(z_{j,A}^{\mathrm{pri},\mathrm{R}},q^{\mathrm{R}}_j)=\int\mathrm{P}(q^{\mathrm{R}}_j|z_{j,A}^{\mathrm{pri},\mathrm{R}},z^{\mathrm{R}}_j)\mathrm{P}(z_{j,A}^{\mathrm{pri},\mathrm{R}},z^{\mathrm{R}}_j)\mathrm{d}z^{\mathrm{R}}_j\\
&=\mathcal{N}{\left(z_{j,A}^{\mathrm{pri},\mathrm{R}};0,\frac{v_x-v_{j,A}^{\mathrm{pri}}}{2}\right)}\int_{l(q^{\mathrm{R}}_j)}^{u(q^{\mathrm{R}}_j)}\int_{-\infty}^{+\infty}
\mathcal{N}{\left(z^{\mathrm{R}}_j;y^{\mathrm{R}}_j,\frac{\sigma^2}{2}\right)} \mathcal{N}{\left(z^{\mathrm{R}}_j;z_{j,A}^{\mathrm{pri},\mathrm{R}},\frac{v_{A}^{\mathrm{pri}}}{2}\right)}\mathrm{d}z^{\mathrm{R}}_j\mathrm{d}y^{\mathrm{R}}_j\\
&\stackrel{(a)}{=}\mathcal{N}{\left(z_{j,A}^{\mathrm{pri},\mathrm{R}};0,\frac{v_x-v_{A}^{\mathrm{pri}}}{2}\right)}\Psi{\left(q^{\mathrm{R}}_j;z_{j,A}^{\mathrm{pri},\mathrm{R}},\frac{\sigma^2+v_{A}^{\mathrm{pri}}}{2}\right)},
\end{aligned}
\end{equation}
where (a) is obtained according to the property given by \cite[(A.7)]{Williams2006} and the definition of $\Psi(\cdot)$. To compute the expectation, we rewrite
$\mathrm{var}{\left[z^R_j\mid q^R_j\right]}$ as follows:
\begin{equation}
\mathrm{var}{\left[z^R_j\mid q^R_j\right]} = \frac{v_A^{\mathrm{pri}}}{2}-{\left(\frac{v_A^{\mathrm{pri}}}{2}\right)}^2\left(\underbrace{\left(\frac{\Psi'(q^{\mathrm{R}}_j;z_{j,A}^{\mathrm{pri},\mathrm{R}},\frac{\sigma^2+v_{A}^{\mathrm{pri}}}{2})}{\Psi(q^{\mathrm{R}}_j;z_{j,A}^{\mathrm{pri},\mathrm{R}},\frac{\sigma^2+v_{A}^{\mathrm{pri}}}{2})}\right)^2}_{ \triangleq{v_1}}-\underbrace{\frac{\Psi''(q^{\mathrm{R}}_j;z_{j,A}^{\mathrm{pri},\mathrm{R}},\frac{\sigma^2+v_{A}^{\mathrm{pri}}}{2})}{\Psi(q^{\mathrm{R}}_j;z_{j,A}^{\mathrm{pri},\mathrm{R}},\frac{\sigma^2+v_{A}^{\mathrm{pri}}}{2})}}_{ \triangleq {v_2}}\right)
\end{equation}
We then compute the expectations of $v_1$ and $v_2$ w.r.t. $(q^{\mathrm{R}}_j,z_{j,A}^{\mathrm{pri},\mathrm{R}})$ as follows:
\begin{subequations}
    \begin{align}
    &\begin{aligned}
    \mathrm{E}[v_1]&=\sum\limits_{b=1}^{2^B}\int_{-\infty}^{\infty}\frac{\left[\Psi'{\left(c_b;z_{j,A}^{\mathrm{pri},\mathrm{R}},\frac{\sigma^2+v_{A}^{\mathrm{pri}}}{2}\right)}\right]^2}{\Psi{\left(c_b;z_{j,A}^{\mathrm{pri},\mathrm{R}},\frac{\sigma^2+v_{A}^{\mathrm{pri}}}{2}\right)}}\mathcal{N}{\left(z_{j,A}^{\mathrm{pri},\mathrm{R}};0,\frac{v_x-v_{A}^{\mathrm{pri}}}{2}\right)}\mathrm{d}z_{j,A}^{\mathrm{pri},\mathrm{R}}\\
    &\stackrel{(b)}{=}\sum\limits_{b=1}^{2^B}\int_{-\infty}^{\infty}\frac{\left[\Psi'{\left(c_b;\sqrt{\frac{v_x-v_{A}^{\mathrm{pri}}}{2}}z,\frac{\sigma^2+v_{A}^{\mathrm{pri}}}{2}\right)}\right]^2}{\Psi{\left(c_b;\sqrt{\frac{v_x-v_{A}^{\mathrm{pri}}}{2}}z,\frac{\sigma^2+v_{A}^{\mathrm{pri}}}{2}\right)}}\mathrm{D}z,
    \end{aligned}\displaybreak[0]\\
    &\begin{aligned}
    \mathrm{E}[v_2]&=\int_{-\infty}^{\infty}\mathcal{N}{\left(z_{j,A}^{\mathrm{pri},\mathrm{R}};0,\frac{v_x-v_{A}^{\mathrm{pri}}}{2}\right)}\sum\limits_{b=1}^{2^B}\Psi''{\left(c_b;z_{j,A}^{\mathrm{pri},\mathrm{R}},\frac{\sigma^2+v_{A}^{\mathrm{pri}}}{2}\right)}\mathrm{d}z_{j,A}^{\mathrm{pri},\mathrm{R}}\\
    &=\int_{-\infty}^{\infty}\mathrm{d}z_{j,A}^{\mathrm{pri},\mathrm{R}}\mathcal{N}{\left(z_{j,A}^{\mathrm{pri},\mathrm{R}};0,\frac{v_x-v_{A}^{\mathrm{pri}}}{2}\right)}\times\\
    &~~~~~~~~~\sum\limits_{b=1}^{2^B}\left(\frac{z_{j,A}^{\mathrm{pri},\mathrm{R}}-r_{b-1}}{\sqrt{\frac{\sigma^2+v_{A}^{\mathrm{pri}}}{2}}}\phi\left(\frac{z_{j,A}^{\mathrm{pri},\mathrm{R}}-r_{b-1}}{\sqrt{\frac{\sigma^2+v_{A}^{\mathrm{pri}}}{2}}}\right)-\frac{z_{j,A}^{\mathrm{pri},\mathrm{R}}-r_{b}}{\sqrt{\frac{\sigma^2+v_{A}^{\mathrm{pri}}}{2}}}\phi\left(\frac{z_{j,A}^{\mathrm{pri},\mathrm{R}}-r_{b}}{\sqrt{\frac{\sigma^2+v_{A}^{\mathrm{pri}}}{2}}}\right)\right)\\
    &\stackrel{(c)}{=}0,
    \end{aligned}
    \end{align}
\end{subequations}
where (b) is obtained by defining the transformation $z_{j,A}^{\mathrm{pri},\mathrm{R}}=\sqrt{\frac{v_x-v_{A}^{\mathrm{pri}}}{2}}z$, and (c) follows from the fact that $\lim_{\eta\to\infty}\eta\phi(\eta)=0$ and $\lim_{\eta\to-\infty}\eta\phi(\eta)=0$. The expectation of $\mathrm{var}{\left[z^\mathrm{I}_j\mid q^\mathrm{I}_j\right]}$ can be computed similarly, and then the expectation of $v^{\mathrm{post}}_{j,A}$ can be obtained by
\begin{equation}\label{e0529_01}
\begin{aligned}
\mathrm{E}{\left[v^{\mathrm{post}}_{j,A}\right]}&=\mathrm{E}{\left[\mathrm{var}[z^\mathrm{R}_j\mid q^\mathrm{R}_j]\right]}+\mathrm{E}{\left[\mathrm{var}[z^\mathrm{I}_j\mid q^\mathrm{I}_j]\right]}\\
&=v_{A}^{\mathrm{pri}}-\frac{(v_A^{\mathrm{pri}})^2}{2}\sum\limits_{b=1}^{2^B}\int_{-\infty}^{\infty}\frac{{\left[\Psi'{\left(c_b;\sqrt{\frac{v_x-v_{A}^{\mathrm{pri}}}{2}}z,\frac{\sigma^2+v_{A}^{\mathrm{pri}}}{2}\right)}\right]}^2}{\Psi{\left(c_b;\sqrt{\frac{v_x-v_{A}^{\mathrm{pri}}}{2}}z,\frac{\sigma^2+v_{A}^{\mathrm{pri}}}{2}\right)}}\mathrm{D}z.
\end{aligned}
\end{equation}
Substituting  \eqref{eIII_08} into \eqref{eA1_04} and \eqref{eA1_05} yields \eqref{eq:SE1} and \eqref{eq:SE2}.

In the same way, $v_{B}^{\rm post}$ converges to the expectation of $v_{j,B}^{\mathrm{post}}$ w.r.t. $x_{j,B}^{\mathrm{pri}}$ and $h'_j$. We first calculate the
expectation of $v_{j,B}^{\mathrm{post}}$ w.r.t. $x_{j,B}^{\mathrm{pri}}$ for the given $h'_j$ elementwisely, i.e., $\mathrm{mmse}(|h'_j|^2\eta)$. Moreover,
substituting $v_{j,B}^{\mathrm{post}}=\mathrm{mmse}(|h'_j|^2\eta)$ into \eqref{e0530_07b} and \eqref{e0411_08a} yields \eqref{eq:SE3}.

\section{Derivation of the Saddle-point of $\mathcal{F}$}\label{AP2}
In this Appendix, we adopt the replica theory in the field of statistical physics to calculate $\mathcal{F}$ in the large-system limit and derive its saddle points, which yield the following proposition.
\begin{Proposition}
    The saddle-point of $\mathcal{F}$ can be obtained from the iteration given by
    \begin{subequations} \label{eq:sdPoint2}
        \begin{align}
            \tilde{q}_{w} &=  v_{x}- \frac{1}{\chi_{s}}, \label{eq:deftqu}\displaybreak[0] \\
            q_{w} &= \frac{1}{2}\sum_{b=1}^{2^{\mathsf{B}}} \int \mathrm{D} v  \frac{\left[\Psi'\left(c_b;\sqrt{\frac{\tilde{q}_{w}}{2}}v,\frac{\sigma^2 + v_{x}-\tilde{q}_{w}}{2} \right)\right]^2}{\Psi\left(c_b;\sqrt{\frac{\tilde{q}_{w}}{2}}v,\frac{\sigma^2 + v_{x}-\tilde{q}_{w}}{2} \right)}, \label{eq:defqu}\displaybreak[0] \\
            \tilde{q}_{s} &= {\left( \frac{1}{ q_{w}} - \frac{1}{\chi_{s}} \right)}^{-1}, \label{eq:deftqx}\displaybreak[0]\\
            \chi_{s} &= \frac{1}{ \frac{1}{N}\sum\limits_{j=1}^{N}|h'_j|^2\mathrm{mmse}(|h'_j|^2\tilde{q}_{s}) } -  \tilde{q}_{s}, \label{eq:defchix}
        \end{align}
    \end{subequations}\hfill\ensuremath{\blacksquare}
\end{Proposition}

\begin{IEEEproof}
    From \cite{Nishimori-01BOOK}, $\mathcal{F}$ can be rewritten as follows:
    \begin{equation}\label{eq:LimF}
        \mathcal{F} = - \frac{1}{N} \lim_{\tau\to 0}\frac{\partial}{\partial \tau}\log \mathrm{E}\left[ \mathrm{P}^{\tau}(\mathbf{q} ;\mathbf{h}')\right].
    \end{equation}
    The expectation operator is moved inside the log-function.
    We first evaluate $\mathrm{E}\left[ \mathrm{P}^{\tau}(\mathbf{q} ;\mathbf{h}')\right]$ for an integer-valued $\tau$, and then generalize the result to any positive real number $\tau$.

    For ease of expression, we denote $\qA = \qF^H\mathrm{diag}(\mathbf{h}) \qP^{\frac{1}{2}}$ and use $\qa_n^H$ to denote the $n$th row of $\qA$.
    Then we rewrite the likelihood \eqref{eII_10} as follows:
    \begin{equation} \label{eq:lnkelihood}
        \mathrm{P}(\mathbf{q} \mid \mathbf{s};\mathbf{h}') \triangleq \prod_{j=1}^{N} \int \rmd  z_{j} \, \mathrm{P}_{\mathrm{out}}(q_j\mid z_j)
        \delta{\left(  z_{j} - \qa_{j}^H \qs \right)},
    \end{equation}
    where $\delta(\cdot)$ denotes Dirac's delta. Using the Fourier representation of the $\delta$  via auxiliary variables $\qw=[ w_m ] \in \bbC^{N}$ to (\ref{eq:lnkelihood}), we obtain
    \begin{equation}\label{eq:lnkelihood2}
        \mathrm{P}(\mathbf{q} ;\mathbf{h}') = {\rm E}_{\qs}\Bigg[ \int \rmd\qz \int \rmd\qw  \, {\mathrm{P}_{\sf out}{\left( \qq \Big| \qz \right)}} e^{-\sfj \qw^H\qz -\sfj\qz^H\qw} \times e^{\sfj \qw^H (\qA \qs) + \sfj (\qA \qs)^H \qw } \Bigg].
    \end{equation}
    Using (\ref{eq:LimF}), we compute the replicate partition function $\mathrm{E}\left[ \mathrm{P}^{\tau}(\mathbf{q} ;\mathbf{h}')\right]$ given by
    \begin{multline} \label{eq:sf_E1_0}
        \mathrm{E}\left[ \mathrm{P}^{\tau}(\mathbf{q} ;\mathbf{h}')\right]
        = \int{ \rmd\qq} ~{\rm E}_{\qA,\qS}\Bigg[ \int{ \rmd\qZ } \int{ \rmd\qW } \\
        \times \left( \prod_{a=1}^{\tau} {\mathrm{P}_{\sf out}{\left( \qq\Big| \qz^{(a)}\right)}} e^{-\sfj \qw^{(a)H} \qz^{(a)} -\sfj \qz^{(a)H} \qw^{(a)} } \right) \times \left( \prod_{a=1}^{\tau} e^{\sfj \qw^{(a) H} \qA \qs^{(a)} + \sfj (\qA \qs^{(a)})^{H} \qw^{(a)} } \right) \Bigg],
    \end{multline}
    where $\qz^{(a)}$ and $\qs^{(a)}$ are the $a$-th replica of $\qz$ and $\qs$, respectively; and $\qZ \triangleq \{
    \qz^{(a)}, \forall a \}$, $\qW \triangleq \{ \qw^{(a)}, \forall a \}$, $\qS \triangleq \{ \qs^{(a)}, \forall a \}$. Here, $\{\qs^{(a)}\}$ are random vectors taken
    from the distribution $\mathrm{P}(\mathbf{s}) $ for $a=1, \dots, \tau$. In addition, $\int\rmd \qq$ denotes the
    integral w.r.t.~a discrete measure because the quantized output $\qq$ is a finite set.

    To evaluate the expectation w.r.t. $\qA$ and $\qS$ in (\ref{eq:sf_E1}), we introduce two $\tau \times \tau$ matrices $\qQ_{s}$ and $\qQ_{w}$ whose elements are
    defined by $[\qQ_{s}]_{a,b}\triangleq\frac{1}{N}\left(\qs^{(a)}\right)^H\qs^{(b)}$ and $[\qQ_{w}]_{a,b}\triangleq\frac{1}{N}\left(\qw^{(a)}\right)^H\qw^{(b)}$. The definitions
    of $\qQ_{s}$ and $\qQ_{w}$ are equivalent to
    \begin{align}
        1 &= \int \prod_{1\leq a \leq b }^{\tau}\delta{\left( {\left(\qs^{(a)}\right)}^H\qs^{(b)} -N[\qQ_{s}]_{a,b}\right)} \rmd  [\qQ_{s}]_{a,b} \\
        1 &= \int \prod_{1\leq a \leq b }^{\tau}\delta{\left( {\left(\qw^{(a)}\right)}^H\qw^{(b)} -N[\qQ_{w}]_{a,b}\right)} \rmd  [\qQ_{w}]_{a,b},
    \end{align}
    where $\delta(\cdot)$ denotes Dirac's delta. Inserting the above expressions into (\ref{eq:sf_E1_0}) yields
    \begin{equation} \label{eq:sf_E1}
        \mathrm{E}\left[ \mathrm{P}^{\tau}(\mathbf{q} ;\mathbf{h}')\right]
        = \int e^{N \calG^{(\tau)}(\qQ_{s},\qQ_{w})} \rmd \mu^{(\tau)}(\qQ_{s}) \rmd \mu^{(\tau)}(\qQ_{w}),
    \end{equation}
    where $\calG^{(\tau)}(\qQ_{s},\qQ_{w})$, $\mu^{(\tau)}(\qQ_{s})$, and $\mu^{(\tau)}(\qQ_{w})$ are given by
    \begin{subequations} \label{eq:defsf_E1}
        \begin{align}
            \calG^{(\tau)}(\qQ_{s},\qQ_{w}) &=  \frac{1}{N} \log {\rm E}_{\qA} {\left[ \prod_{a=1}^{\tau} e^{-\sfj \qw^{(a) H} \qA \qs^{(a)} - \sfj (\qA \qs^{(a)})^{H}\qw^{(a)}  } \right]}, \label{eq:defG} \displaybreak[0]\\
            \mu^{(\tau)}(\qQ_{s}) &= {\rm E}_{\qS}{\left[\int \prod_{1\leq a \leq b }^{\tau}\delta{\left( {\left(\qs^{(a)}\right)}^H\qs^{(b)} -N[\qQ_{s}]_{a,b}\right)} \rmd  [\qQ_{s}]_{a,b} \right]}, \label{eq:defmux} \displaybreak[0]\\
            \mu^{(\tau)}(\qQ_{w}) &= \int{ \rmd\qq} \int{ \rmd\qZ } \int{ \rmd\qW }
            {\left( \int \prod_{1\leq a \leq b }^{\tau}\delta{\left( {\left(\qw^{(a)}\right)}^H\qw^{(b)} -N[\qQ_{w}]_{a,b}\right)} \rmd  [\qQ_{w}]_{a,b} \right)} \nonumber \displaybreak[0]\\
            & \hspace{1cm} \times \prod_{a=1}^{\tau} {\mathrm{P}_{\sf out}{\left( \qq\mid \qz^{(a)}\right)}} e^{-\sfj \qw^{(a)H} \qz^{(a)} -\sfj \qz^{(a)H} \qw^{(a)} }. \label{eq:defmuu}
        \end{align}
    \end{subequations}
    We notice that by introducing the $\delta$-functions, the expectations over $\qS$ can be separated into an expectation over all possible
    covariance $\qQ_{s}$ and all possible $\qS$ configurations w.r.t. a prescribed set of $\qQ_{s}$. Therefore, we can separate the expectations over $\qA$ and $\qS$ respectively in (\ref{eq:defG}) and (\ref{eq:defmux}).
    A similar concept applies to separating the expectations over $\qA$ and $\qW$.
    We next calculate each term of (\ref{eq:defsf_E1}).

    First, we evaluate $\calG^{\tau}(\qQ_{s},\qQ_{w})$ by noticing 
    \begin{equation} \label{eq:E_uAx1}
    {\rm E}_{\qA} {\left[ \prod_{a=1}^{\tau} e^{-\sfj \qw^{(a) H} \qA \qs^{(a)} - \sfj (\qA \qs^{(a)})^{H}\qw^{(a)}  } \right]}
    = {\rm E}_{\qA} {\left[ e^{-\sfj \sum_{a=1}^{n} \tqw^{(a)H} \qLambda \tqs^{(a)} + \tqs^{(a)H} \qLambda \tqw^{(a)} } \right]},
    \end{equation}
    where $\qLambda^{\frac{1}{2}} = \mathrm{diag}(\mathbf{h}) \qP^{\frac{1}{2}}$, $\tqw^{(a)} = \qF \qw^{(a)} $, and $\tqs^{(a)} =  \qLambda^{\frac{1}{2}} \qs^{(a)}$. The covariances of $(\tqs^{(a)},\tqs^{(b)})$ and $(\tqw^{(a)},\tqw^{(b)})$ are given by the following:
    \begin{align}
    &\frac{1}{N}\left(\tqs^{(a)}\right)^H\tqs^{(b)} = \frac{1}{N}\left(\qs^{(a)}\right)^H\qLambda\qs^{(b)} = [\qQ_{s}]_{a,b}, \label{eq:defQx}\\
    &\frac{1}{N}\left(\tqw^{(a)}\right)^H\tqw^{(b)} = \frac{1}{N}\left(\qw^{(a)}\right)^H\qw^{(b)} = [\qQ_{w}]_{a,b}. \label{eq:defQu}
    \end{align}
    Notice that the dependence on the replica indices would not affect the physics of the system because replicas have been introduced artificially. Assuming {\it replica symmetry} (RS), i.e.,
    \begin{equation}
    \left\{\begin{aligned}
    \qQ_{s}&=c_{s} \qI_{\tau}+q_{s}{\bf 11}^H,\\
    \qQ_{w}&=c_{w} \qI_{\tau}-q_{w}{\bf 11}^H,
    \end{aligned}\right.
    \end{equation}
    therefore seems natural.
    With the RS, we can obtain follows \cite{Shinzato-09JPA}:
    \begin{equation} \label{eq:E_uAx2}
    \calG^{(\tau)}(\qQ_{s},\qQ_{w})
    =  (\tau-1)G(c_{s},c_{w}) + G(c_{s}+\tau q_{s},c_{w}-\tau q_{w}),
    \end{equation}
    where
    \begin{equation} \label{eq:E_uAx3}
    G(x,w)
    =  \Extr_{\chi_{s},\chi_{w}} \Big\{ \chi_{s} x + \chi_{w} w
    -  \log(\chi_{s}\chi_{w} + 1 ) \Big\}
    - \log x -\log w - 2,
    \end{equation}
    and $\Extr_{x}\{ f(x) \}$ denotes the extreme value of $f(x)$ w.r.t.~$x$.

    Next, we consider $\mu^{(\tau)}(\qQ_{s})$ in (\ref{eq:defmux}). It can be shown that $\mu^{(\tau)}(\qQ_{s}) = e^{N \calR_{s}^{(\tau)}(\qQ_{s})+\calO(1)}$, where
    $\calR_{s}^{(\tau)}(\qQ_{s})$ is the rate measure of $\mu^{(\tau)}(\qQ_{s})$ and is given by \cite{Wen-07IT}
    \begin{equation} \label{eq:Rx1}
        \calR_{s}^{(\tau)}(\qQ_{s}) = \max_{\tqQ_{s}}\left\{
        \frac{1}{N}\log{\rm E}_{\tqS}{\left\{e^{\tr\left(\tilde\qQ_{s} \tqS^H\tqS\right)}\right\}}
        - \tr{\left(\tqQ_{s} \qQ_{s}\right)} \right\}
    \end{equation}
    with $\tilde\qQ_{s}\in {\mathbb R}^{\tau\times\tau}$ being a symmetric matrix.
    Furthermore, we assume the RS, i.e., $\tqQ_{s} =  \tq_{s} {\bf 11}^H - \tc_{s} \qI_{\tau}$.
    With the RS, and using the Hubbard-Stratonovich transformation and introducing the auxiliary vector
    $\qu_{s} \in \mathbb{C}^{N}$, the first term of (\ref{eq:Rx1}) can be written as follows:
    \begin{equation}\label{eq:LT_Rx2}
    \small
        \frac{1}{N}\log{\rm E}_{\tqS}{\left[e^{\tr\left(\tilde\qQ_{s} \tqS^H\tqS\right)}\right]}=\int \rmd\qu_{s}
        \left( {\rm E}_{\tqs}{\left[e^{ -\| \qu_{s} - \sqrt{\tq_{s}}\tqs\|^2 + (\tq_{s}-\tc_{s})\tqs^{H}\tqs }\right]}\right)
        \left( {\rm E}_{\tqs}{\left[e^{\left(\sqrt{\tq_{s}}\tqs\right)^H \qu_{s}+\qu_{s}^H\sqrt{\tq_{s}}\tqs-\tc_{s}\tqs^H \tqs}\right]} \right)^{\tau-1}.
    \end{equation}
    With the RS assumption, the last term of (\ref{eq:Rx1}) can now be expressed as follows:
    \begin{equation} \label{eq:RSQQ}
    \tr{\left(\tilde\qQ_{s}\qQ_{s}\right)}
    = (-\tc_{s}+\tau\tq_{s})(c_{s}+\tau q_{s}) - (\tau-1) \tc_{s}c_{s}.
    \end{equation}
    Substituting (\ref{eq:LT_Rx2}) and (\ref{eq:RSQQ}) into (\ref{eq:Rx1}) and taking the derivative w.r.t. $\tau$ at $\tau = 0$, we obtain the following:
    \begin{multline} \label{eq:FinaRx}
        \left. \partial\calR_{s}^{(\tau)}(\qQ_{s})/\partial\tau \right|_{\tau=0} = \max_{\tc_{s},\tq_{s}}\Bigg\{
        \int \rmD u_{s} {\rm E}_{\ts}{\left[ e^{- |u_{s}-\sqrt{\tq_{s}}\ts|^2 + (\tq_{s}-\tc_{s}) |\ts|^2 }   \right]}\\
        \times \log{\rm E}_{s}{\left[ e^{-\tc_{s}|\ts|^2 + {\rm Re}\left[\sqrt{\tq_{s}} u_{s}^{*} \ts \right] }   \right]}
        - \tc_{s} (c_{s}+q_{s}) + \tilde{q}_{s} c_{s} \Bigg\}.
    \end{multline}

    Similarly, we calculate $\mu^{(\tau)}(\qQ_{w})$ in (\ref{eq:defmuu}) and assume the RS $\tqQ_{w} =  -\tq_{w} {\bf 11}^H - \tc_{w} \qI_{\tau}$. It can be shown that
    $\mu^{(\tau)}(\qQ_{w}) = e^{N \calR_{w}^{(\tau)}(\qQ_{w})+\calO(1)}$, where $\calR_{w}^{(\tau)}(\qQ_{w})$ is the rate measure of $\mu^{(\tau)}(\qQ_{w})$ and is
    given by the following:
    \begin{equation} \label{eq:Rw1}
    \calR_{w}^{(\tau)}(\qQ_{w})= \max_{\tqQ_{w}}\left\{
    \frac{1}{N}\log\calR_{w1}(\tilde\qQ_{w})
    - \tr\left(\tqQ_{w} \qQ_{w}\right) \right\},
    \end{equation}
    where we define
    \begin{equation}
    \calR_{w1}(\tilde\qQ_{w}) \triangleq \int{ \rmd\qq } \int{ \rmd\qZ } \int{ \rmd\qW } \Bigg(
    \prod_{a=1}^{\tau} {\mathrm{P}_{\sf out}{\left( \qq \mid \qz^{(a)}\right)}}
    \times e^{-\sfj \qw^{(a)H} \qz^{(a)} -\sfj \qz^{(a)H} \qw^{(a)} } \Bigg)
    e^{\tr\left(\tilde\qQ_{w} \qW^H\qW\right)}.
    \end{equation}
    By using the Hubbard-Stratonovich transformation and introducing the auxiliary
    vector $\qu_{w} \in \mathbb{C}^{N}$, we obtain
    \begin{equation}
    \begin{aligned}\label{eq:Rw2}
     \calR_{w1}(\tilde\qQ_{w})&= \int \rmD\qu_{w} \Bigg( \prod_{a=1}^{\tau} \int \rmd\qz^{(a)} \int \rmd\qw^{(a)} {\mathrm{P}_{\sf out}{\left( \qq \mid \qz^{(a)}\right)}}\times e^{-\sfj \qw^{(a)H} \qz^{(a)} -\sfj \qz^{(a)H} \qw^{(a)} } \Bigg) \nonumber \\
    & ~~ \times  e^{ (\sum_{a} \sfj \sqrt{\tq_{w}}\qw^{(a)})^{H} \qu_{w}
        + \qu_{w}^{H} (\sum_{a} \sfj \sqrt{\tq_{w}}\qw^{(a)})  - \sum_{a}\tc_{w}\qw^{(a)H} \qw^{(a)} }\\
    &= \int \rmD\qu_{w} \left( \int \rmD\qv_{w} \, {\mathrm{P}_{\sf out}{\left( \qq \Big| \sqrt{\tc_{w}} \qv_{w} + \sqrt{\tq_{w}}\qu_{w} \right)}}  \right)^{\tau},
    \end{aligned}
    \end{equation}
    where the last equality follows the facts that $\qv_{w} \triangleq \frac{1}{\sqrt{\tc_{w}}} \left(\sqrt{\tq_{w}}\qu_{w} - \qz \right)$ and $\rmD\qv_{w} = \frac{1}{\pi^{N}} e^{-\qv_{w}^H\qv_{w}} $.
    With the RS assumption, the last term of (\ref{eq:Rw1}) can now be expressed as follows:
    \begin{equation} \label{eq:RSQQw}
    \tr{\left(\tilde\qQ_{w}\qQ_{w}\right)}
    = (-\tc_{w}+\tau\tq_{w})(c_{w}+\tau q_{w}) - (\tau-1) \tc_{w}c_{w}.
    \end{equation}
    Substituting (\ref{eq:Rw2}) and (\ref{eq:RSQQw}) into (\ref{eq:Rw1}) and taking the derivative w.r.t. $\tau$ at $\tau = 0$, we obtain the following:
    \begin{multline} \label{eq:FinaRw}
        \left. \partial\calR_{w}^{(\tau)}(\qQ_{w}) /\partial\tau \right|_{\tau=0} = \max_{\tc_{w},\tq_{w}}\Bigg\{
        \sum_{q} \, \rmD u_{w} {\left( \int \rmD v_{w} \mathrm{P}_{\sf out}{\left( q \Big| \sqrt{\tc_{w}} v_{w} + \sqrt{ \tilde{q}_{w}} u_{w}  \right)} \right)}\\
        \times \log   {\left( \int \rmD v_{w} \mathrm{P}_{\sf out}{\left( q \Big| \sqrt{\tc_{w}} v_{w} + \sqrt{ \tilde{q}_{w}} u_{w}  \right)} \right)} - \tc_{w} (q_{w}+c_{w}) + \tilde{q}_{w} c_{w}
        \Bigg\}.
    \end{multline}

    Applying (\ref{eq:Rx1}) and (\ref{eq:Rw1}) the integration over $\qQ$ in (\ref{eq:sf_E1}) can be performed via the saddle point method as $N\rightarrow\infty$,
    which yields the following:
    \begin{equation}\label{eq:sadP}
    \lim_{N\rightarrow\infty}\frac{1}{N} \mathrm{E}{\left[ \mathrm{P}^{\tau}(\mathbf{q} ;\mathbf{h}')\right]} \\
    = \max_{\qQ_{s},\qQ_{w}}\Big\{{\cal G}^{(\tau)}(\qQ_{s},\qQ_{w})-
    \calR_{s}^{(\tau)}(\qQ_{s}) - \calR_{w}^{(\tau)}(\qQ_{w}) \Big\} \triangleq -\calF^{(\tau)} .
    \end{equation}
    With the normalization constraint $ \mathrm{E}\left[ \mathrm{P}^{\tau}(\mathbf{q} ;\mathbf{h}')\right] = 1$, we can obtain that $c_{s} + q_{s} =  v_{x} $, $c_{w} - q_{w} = 0$, $-\tc_{s} + \tq_{s} = 0$, and $\tc_{w} + \tq_{w} = v_{x}$. Substituting (\ref{eq:E_uAx2}), (\ref{eq:FinaRx}), and (\ref{eq:FinaRw}) into (\ref{eq:sadP}), and combining it with the aforementioned relationships, we obtain $\partial\calF^{(\tau)}/\partial\tau$ at $\tau=0$ as follows:
    \begin{multline}\label{eq:GenFree2}
    \calF = \Extr_{q_{s},q_{w}}\Bigg\{ G(v_{x}-q_{s},q_{w})  + q_{w} v_{x} - \frac{1}{N} \sum_{j=1}^{N} I\left(x_j;\ty_j \left|\sqrt{\tq_{s}}h'_{j} \right.\right)\\
    + \tq_{s} (v_{x}-q_{s}) + \sum_{q} \int \rmD v \calP_{\sf out}(q| v; \tq_{w}) \log \calP_{\sf out}(q| v; \tq_{w})
    - \tilde{q}_{w} q_{w} \Bigg\},
    \end{multline}
    where
    \begin{align}
    &\calP_{\sf out}(q| v; \tq_{w}) = \int \rmD u \mathrm{P}_{\sf out}{\left( q \Big| \sqrt{ v_{x} -\tq_{w}} u + \sqrt{ \tilde{q}_{w}} v  \right)},\\
    &I{\left(s_{j};\tilde{y}_{j}\left|\sqrt{\tilde{q}_{x}} h'_{j} \right.\right)}
    = -  \int \mathrm{d} \tilde{y}_{j}  \mathcal{P}_{s}(\tilde{y}_{j})
    \log \mathcal{P}_{s}(\tilde{y}_{j}) - 1,\\
    &\mathcal{P}_{s}(\tilde{y}_{j}) = \int \mathrm{d} s_{j} \mathrm{P}(s_{j})\frac{1}{\pi} e^{-\left| \tilde{y}_{j} - \sqrt{\tilde{q}_{x}} h'_js_j \right|^2}.
    \end{align}
    The saddle-point of (\ref{eq:GenFree2}) can be rewritten as
    \begin{subequations} \label{eq:sdPoint2}
        \begin{align}
        \tq_{w} &= v_{x}+ \chi_{w} - \frac{1}{q_{w}}, \label{eq:sdPoint2_1}\displaybreak[0]\\
        q_{w} &= \frac{1}{2}\sum_{b=1}^{2^{\mathsf{B}}} \int \mathrm{D} v  \frac{\left[\Psi'\left(c_b;\sqrt{\frac{\tilde{q}_{w}}{2}}v,\frac{\sigma^2 + v_x - \tilde{q}_{w}}{2} \right)\right]^2}{\Psi\left(c_b;\sqrt{\frac{\tilde{q}_{w}}{2}}v,\frac{\sigma^2 + v_x -\tilde{q}_{w}}{2} \right)}, \label{eq:sdPoint2_2} \displaybreak[0]\\
        \tq_{s} &= -\chi_{s} + \frac{1}{v_{x}-q_{s}}, \label{eq:sdPoint2_3} \displaybreak[0]\\
        v_{x}-q_{s} &= \frac{1}{N}\sum\limits_{j=1}^{N}|h'_j|^2\mathrm{mmse}(|h'_j|^2\tilde{q}_{s}) \label{eq:sdPoint2_4}
        \end{align}
    \end{subequations}
    From (\ref{eq:E_uAx2}), we obtain that the extremum points should satisfy the following equality
    \begin{equation}
    q_{w} = \frac{\chi_{s}}{\chi_{s}\chi_{w}+1},~~~
    v_{x}-q_{s} = \frac{\chi_{w}}{\chi_{s}\chi_{w}+1}. \label{eq:qu-Tx-qx}
    \end{equation}
    Substituting (\ref{eq:qu-Tx-qx}) into (\ref{eq:sdPoint2_1}), (\ref{eq:sdPoint2_3}) and \eqref{eq:sdPoint2_4}, we obtain Proposition 2.
\end{IEEEproof}

\ifCLASSOPTIONcaptionsoff
  \newpage
\fi

\end{document}